# Droplet nanofluidic transport under vapor deposition: a review on seeded growth of low-dimensional nanomaterials

Zheng Fan, Lei Ma

TICNN

Abstract

Thin film deposition technologies boost the development of modern semiconductor industries. Being a fancy variant, vapor phase deposition on metal nanoparticles (often in liquid phase) rather than on bare substrates opens novel avenues of fabricating low-dimensional nanomaterials, which renders the development of new device architectures and their applications in advanced electronics, optoelectronics and photonics, etc. Since the last twenty years, nanomaterials with various geometries (i.e. dots, wires, trees, tubes, flakes, ribbons, etc.) have been synthesized via different bottom-up methods (i.e. vapor-liquid-solid, vapor-solid-solid, in plane solid-liquid-solid, etc.) by different deposition techniques (CVD, PECVD, MOCVD, MBE, etc.). In contrast with liquid phase epitaxy where metal liquid severs as stationary reservoir that accommodates gaseous precursors, metal droplets have to be kicked off in-plane on/out-of-plane from the substrates so as to steer the growth of low-dimensional nanomaterials. In this review, we shall regard the growth process in a viewpoint of dynamic droplet evolution under vapor phase deposition. We shall summarize several key factors that affect the droplet spreading behaviors and their consequent nanofluidic transport, which involves deposition parameters, solid-liquid interfaces, crystal phases, substrate nanofacets and so on, which deterministically results in various morphologies and growth directions of the nanomaterials. Reversely, the aspects like doping profile and phase transition that are strongly dependent on the droplet transport will also be discussed.

1. Introduction

Thanks to their unique properties like high specific surface area and aspect ratio, rich surface and edge states, and flexible combination for heterostructures and superlattices, low-dimensional nanomaterials allows the realization of novel device architectures in vast applications and provides new platforms for fundamental studies like quantum physics.[1-29] In contrast to the top-down methods where nanostructures are tailored from bulk materials by advanced lithography technologies, the growth of low-dimensional nanomaterials, like 0-D quantum dots (QDs), 1-D semiconductor nanowires (NWs), carbon nanotubes (CNTs) or even nanoribbons (NRs) of 2-D materials, paves promising routes for achieving smooth surfaces, high quality interfaces and well-arranged edges in atomic scale. Analogue with their 3-D counterpart of thin film growth, the syntheses of low-dimensional nanomaterials often rely on vapor deposition, where gaseous precursors are accommodated within metal droplets rather than within liquid metal pool or directly stick and diffuse on bare solid surfaces. As it is the metal droplets that mediate the nanomaterials growth, the transport behaviors of the droplets must predominantly determine the growth processes and geometries of the nanomaterials, either they may spread to form dots, flakes, dendrites, or they may crawl horizontally or be lifted up out-of-plane to produce wires or tubes. Moreover, the interfaces between the droplets and the nanomaterials also affects the properties of the nanomaterials like crystalline phases and doping profiles, which offers opportunities of their controllable engineering. On the other hand, as the droplet transport runs in an environment where vapor deposition proceeds, this thereby enables fine tuning of the deposition parameters to design the nanomaterials deterministically. In this review, we shall regard the low-dimensional nanomaterials growth from a viewpoint of droplet nanofluidic transport. By starting from the principles of droplet spreading and spontaneous motion, we shall summarize several representative droplet nanofluidic phenomena under vapor deposition, and attempt to discuss the effects of the deposition parameters and the substrate morphologies on the geometries of nanomaterials and on their physical and chemical properties.

2. Principles of droplet wetting and spontaneous motion on solid surface

Prior to discussing the naonfluidic transport phenomena in metal droplet mediated low-dimensional nanomaterials growth, we shall introduce the principles of droplet wetting and spontaneous motion on solid surface, which have been substantially studied and employed in many

applications.[30-63]

Consider a solid (S)-liquid (L)-vapor (V) system where a droplet is placed on a solid surface in a certain gaseous atmosphere, the droplet may cover the solid surface by forming a interface with a contact angle $\theta$ for the sake of total surface energy minimization, which is called wetting.[30, 64] In this review, we only consider the droplets smaller than their Laplace lengths $\kappa^{-1}$ so that the gravitational effect can be neglected, otherwise larger ones will shape like "pancakes".[43] As illustrated in Figure 1 (a-1), the droplet forms into a spherical cap on the solid surface. The thermodynamic equilibrium condition of this S-L-V system is expressed by Young's equation:

$$\gamma_{SV} - \gamma_{SL} - \cos\theta \cdot \gamma_{LV} = 0, (1)$$

where $\theta$ denotes the S-L contact angle in equilibrium state, $\gamma_{SV,SL,LV}$ denote solid surface energy, solid-liquid interfacial energy and droplet surface energy, respectively.[30, 65] Note that surface or interface tensions are also used in literatures, no matter for solid or liquid phases. However, if the droplet reacts with the solid surface with the production of a new phase at the S-L interface, its wetting behavior may keep altering until the system approaches a new state of thermodynamic equilibrium, which often appears in systems of liquid metals on solid metal, ceramic or semiconductor surfaces.[65-75] The Reaction Product Control (RPC) model is the most popular one that describes the reactive wetting behaviors [41, 65, 71]: as illustrated in Figure 1 (a-2), the chemical reaction of producing new compound P at the S-L interface causes the loss of the system's Gibbs free energy ($\Delta G_R$) which is completely attributed to the decrease of the interfacial energy component

$$\Delta\gamma_{SL} = \Delta G_R/A, (2)$$

where $\Delta\gamma_{SL}$ denotes the S-L interfacial energy loss per unit area (with negative sign) and $A$ the S-L contact area. As a consequence, the balance of Young's equation in Figure (a-1) breaks:

$$\gamma_{SV} - \left(\gamma_{SL} + \frac{\Delta G_R}{A}\right) - \cos\theta \cdot \gamma_{LV} > 0, (3)$$

thereafter the droplet starts to spread and the S-L-V triple phase line (TPL) advances to the unreacted solid surface with decreasing contact angle until $\theta_{min}$, then relaxes up to a new equilibrium state once the new phase P at the S-L interface thoroughly impedes the chemical reaction, as seen in Figure 1 (a-3). This is due to the fact that chemical reaction often proceeds fast at its initial stage and then slows down until the entire process terminates. The corresponding time dependent interfacial energy profile is schematically plotted in Figure 1 (b), where $\gamma\alpha\beta$ represents the interfacial energy (tension used in the figure) of two phases $\alpha$, $\beta$, and $\Delta g^{\alpha\beta}$ equals $\Delta\gamma_{SL}$ in Equation (2).[64] The possible processes of atom diffusion, nucleation and crystal growth at the triple phase line (TPL) are illustrated in Figure 1 (c), by taking liquid Pb-Sn alloy on Cu surface for instance. And an SEM image of Sn droplet reactively wetting Cu surface with intermetallic $Cu_6Sn_5$ formed at the interface is shown in Figure 1 (d).

Since the balance of the Young's equation can be broken by a dynamic S-L interface, it makes possible the droplet motion by preparing chemically inhomogeneous solid surfaces. The principle of driving a droplet transport is to establish and maintain a solid surface wettability gradient. Let us consider a one-step motion of a droplet on a solid surface with asymmetric surface energy at its two ends, as illustrated in Figure (e): (1) at initial stage just before moving, the droplet follows the Young's equation at both ends with different contact angles $\theta_{A,B}$:

$$\gamma_{SV(A,B)} - \gamma_{SL(A,B)} - \cos\theta_{A,B} \cdot \gamma_{LV} = 0, \quad (4)$$

where $\gamma_{SV(A)}$ is larger than $\gamma_{SV(B)}$; (2) however the Laplace pressure within the droplet will correct such distorted droplet into a spherical cap with isotropic curvatures (in a 2D viewpoint) and thereby the droplet acquires identical dynamic contact angles $\theta_d$ ($\theta_A<\theta_d<\theta_B$) at its both ends. As a consequence, the Young's equation is unbalanced with a force $dF$ (per unit area) pushing the droplet to the high surface energy side:

$$\begin{aligned} dF &= [\gamma_{SV(A)} - \gamma_{SL(A)} - \cos\theta_d \cdot \gamma_{LV}] - [\gamma_{SV(B)} - \gamma_{SL(B)} - \cos\theta_d \cdot \gamma_{LV}] \\ &= [\gamma_{SV(A)} - \gamma_{SL(A)}] - [\gamma_{SV(B)} - \gamma_{SL(B)}] = \gamma_{LV}(\cos\theta_A - \cos\theta_B) \end{aligned}. \quad (5)$$

From the thermodynamic viewpoint, the droplet tends to cover and thus reduce the higher surface energy area so that to minimize the system total Gibbs free energy.

Various strategies have been demonstrated to realize droplet self-propulsions, by building up wettability gradients on solid surfaces via direct chemical treatments of solid surfaces or relying on auto-hydrophobicity/hydrophilicity. As seen in Figure 1 (f-1), the solid surface turns to be hydrophobic after silanization treatment, however the droplet dissolves the silanes so that recover the hydrophilicity of the solid surface hydrophilic, which results in its free running. Such auto-hydrophilic process possesses a self-avoiding feature where the droplet never crosses its moving trail (see Figure 1 (f-2)).[46] Moreover, its random trajectory can be regulated by designing the direction of solid surface energy gradient (see Figure 1 (f-3)), or even upwards along an inclined surface (see Figure 1 (h)).[47] Another case worth presenting here is the *in situ* LEEM observation of tiny 2-D Sn islands sweeping the Cu surface, leaving bronze (i.e. Sn-Cu intermetallic) in its track, where the author also attributed this migration to the alloying induced surface energy lowering.[50, 76] In general, droplet self-propulsions often rely on S-L interactions to change solid surface energies as well as interfacial energies, either via dissolution of solid surface elements like that in Figure 1 (f), or via reactions like that in Figure 1 (g). Such behaviors of droplet nanofluidic transport are in principle attributed to the droplet spontaneous motion induced by reactive wetting, which has been theoretically elaborated by de Gennes.[44]

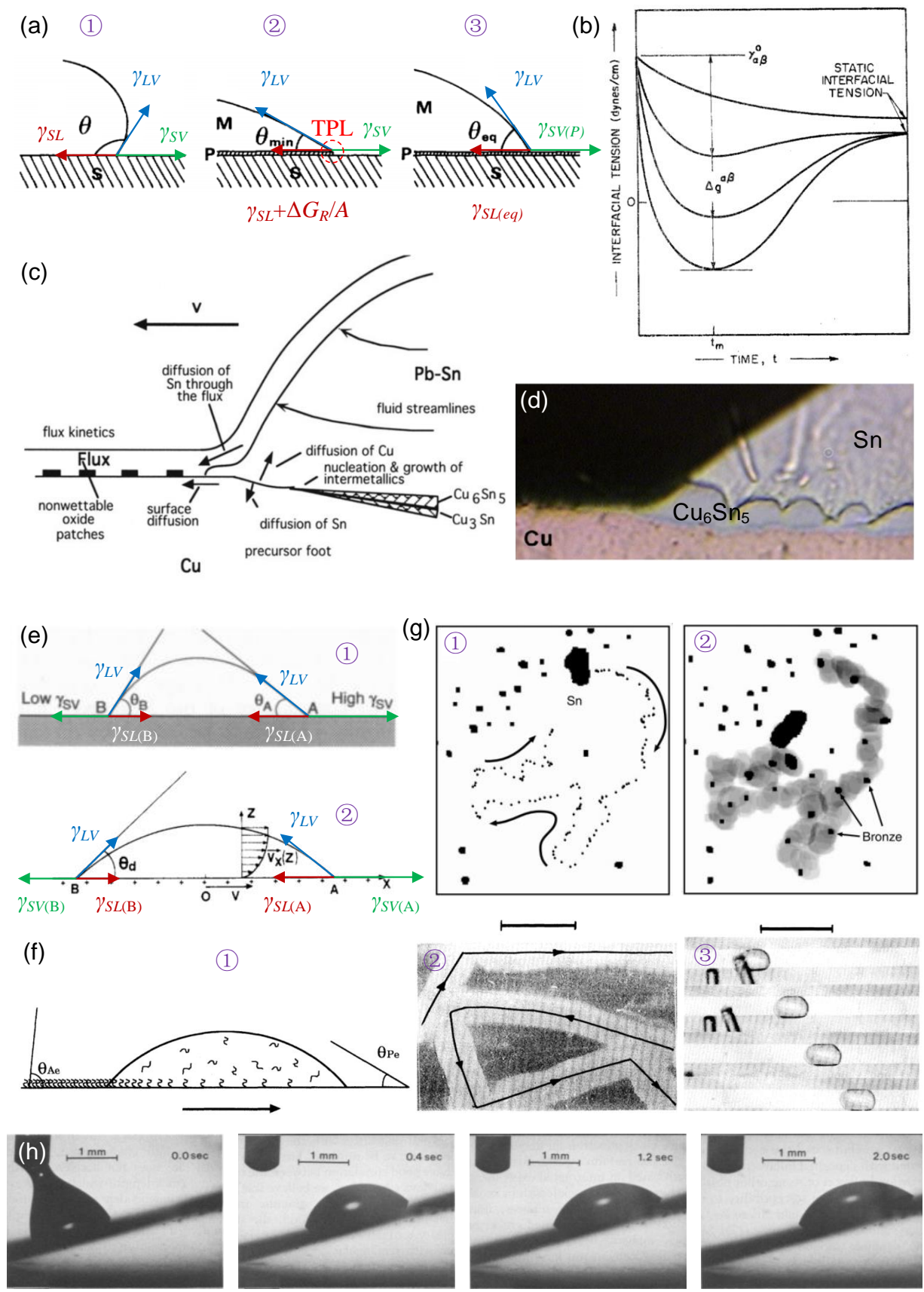

Figure 1. Summary of droplet wetting and spontaneous motion on solid surface. (a) Schematic

representation of a droplet placed on a solid surface: (1) non-reactive wetting, (2) transient stage of reactive wetting with new product P formed at the S-L interface, and (3) new equilibrium stage after the new product P completely impedes the chemical reaction. (b) Evolution of dynamic interfacial energy between two phases during a chemical reaction. (c) Schematic representation of possible processes occurring during a Cu substrate reactively wetted by a Pb-Sn droplet. (d) SEM images of a Sn droplet wetting on Au with intermetallic $Cu_6Sn_5$ formed at the interface at 250 °C for 295 s. (e) Schematic representations of (1) asymmetric wetting of a droplet on a chemically inhomogeneous solid surface with different contact angles at the two ends, and of (2) transient contact angle of a moving droplet on a solid surface. (f) (1) Schematic representation of a self-propelling droplet induced by auto-hydrophilicity: the silanization by a liquid membrane makes the solid surface hydrophobic (see the left end of the droplet), once the droplet dissolves the silanes and thereby turns the solid surface hydrophilic (see the right end), its migration is activated. Such a spontaneous motion possesses (2) a special signature of self-avoiding behavior and (3) its trajectory can be designed by patterning the solid surface wettability artificially. (g) *In situ* LEEM observation shows that tiny 2-D Sn islands alloys with the Cu substrate and thereafter lowers the substrate surface energy, which renders the following migration of Sn islands with a bronze trail behind. (h) A sequence of optical images of a water droplet running uphill on an inclined solid surface with surface energy gradient.

## 3. Droplet epitaxy of quantum dots/rings and dendrites

In a general picture under vapor phase deposition, the 1-D nanomaterials are normally produced once droplets are activated to move, while the 0-D ones are obtained by stationary droplets. Following the brief introduce of the principles of droplet wetting and spontaneous motion, we shall start from the stationary droplets in this section, which will tell us how they get pinned in bulk at their original spots and can be regarded as the prelude of the mobile droplets for 1-D nanomaterials growth.

Vapor phase epitaxy of quantum dots, of III-V, III-Nitride or IV group materials, are commonly achieved via the Stranski-Krastanow (SK) growth mode, where the surface reconstruction benefits from the increasing strain and the consequent lattice relaxation at the lattice-mismatched QD/buffer layer interfaces.[77-81] Derived from the SK technology, the droplet epitaxy method provides broad opportunities to engineer the morphologies of the quantum nanostrucutres by fine-tuning the deposition parameters involving the substrate temperature and the atom flux beam equivalent pressure (BEP).[82-96] Figure 2 (a) depicts a typical droplet epitaxy process by taking GaAs QDs for instance, where a Ga droplet is predeposited on the AlGaAs buffer layer, followed by the As flux irradiation, whose atoms will adsorb and diffuse either directly onto the droplet (i.e. region I), either on the substrate surface away from the droplet (region III), or on the region II which is in the range of the Ga diffusion.[92] The thermodynamic calculation demonstrates that GaAs nucleus prefers to precipitate at the TPL (i.e. position B in Figure 2 (b)) with the lowest Gibbs free energy cost, which is consistent with the 1-D VLS growth process.[90, 97-99] Furthermore, by varying the substrate temperature and As flux BEP (see Figure 2 (c)), it renders the shape evolution from quantum dots to quantum rings, double rings or even quantum holes, as illustrated in Figure 2 (d).[90]

For an epitaxial growth process via SK mode as illustrated in Figure 2 (g), obtaining a QD rather than a continuous wetting layer requires a thermodynamic status as below:

$$E_{epilayer} + E_{strain} > E_{surface}, (6)$$

where $E_{epilayer}$, $E_{strain}$ and $E_{surface}$ denote the Gibbs free energies of the epitaxial layer (i.e. epilayer), the strain arising from the lattice mismatch, and the substrate surface (i.e. buffer layer here), respectively. In order to meet the criterion above, lowering the substrate surface energy is a proper route. In principle, the effect of As amount on the substrate surface energy can be expressed by

$$\Delta E_{surface} = (N_{Ga} - N_{As})\Delta\mu_{As}, (7)$$

where $\Delta E_{surface}$ denotes the substrate surface energy change, $\Delta\mu_{As}$ denotes the As chemical potential change, $N_{Ga}$ and $N_{As}$ denote the densities of Ga and As atoms, repectively. This can be realized by introducing high As flux, meanwhile the As effusion can also be suppressed at low substrate temperature.[100, 101]

Likewise, in a droplet epitaxy process, droplet dewetting will yield a QD while a quantum ring will be obtained once droplet wets the substrate, which is in agreement with the schemes in Figure 2 (d). Moreover, besides rendering an As rich substrate surface with low free energy, low temperature also impedes the Ga atomic diffusion, while high As flux accelerates the crystallization of Ga droplet into GaAs.[90, 93, 96, 102] Reversely, high temperature and low As flux allows the droplet gradually spreading on the substrate which results in rings or holes. The above criteria are well met by the MBE experiments, as shown in Figure 2 (e, f) by fixing the substrate temperature and tuning the As flux BEP or vice versa.[94, 95] In comparison with VLS process of III-V or III-Nitride NWs growth, the main difference is that the droplet epitaxy proceeds by exposing V group or N flux to Ga or In droplets, with no supplement of corresponding droplet atoms, thus instead of being lifted up, the droplets have to be fixed at their original spots until their complete consumption by crystallization, no matter they wet or dewet the substrates.

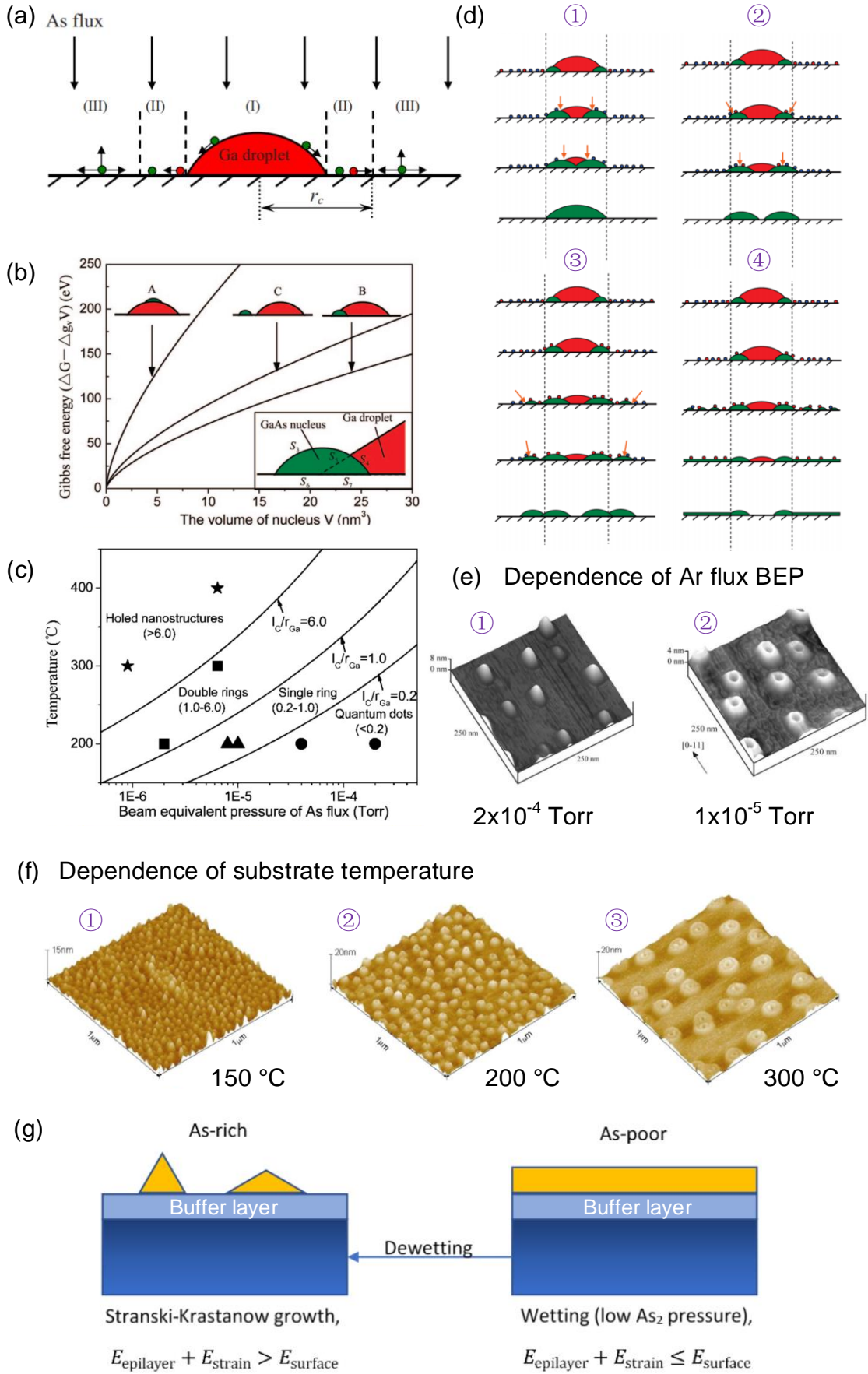

Figure 2. Evolution of III-V nanostructures from quantum dots to quantum rings via droplet epitaxy. (a) Schematic representation of the kinetic process of Ga droplet under As flux exposure. The arrows represent the possible surface diffusion process of As (green) and Ga (red) atoms. Region (I) represents the Ar deposition on Ga droplet, region (II) and (III) represents the As deposition on the areas within/out of the range of Ga atom surface diffusion, respectively. (b) Gibbs free energy difference as a function of the GaAs nucleus volume at three different sites, which indicates that the nucleation at the TPL (i.e. position B) consumes the lowest energy. (c) Effects of substrate temperature and As flux BEP on the morphologies of the GaAs quantum nanostructures via theoretical calculations. (d) Schematic representation of the evolution of the morphologies of the GaAs nanostructures with different degrees of Ga atom surface diffusion, from (1) quantum dots, (2) quantum rings, (3) double rings to (4) holes. The vertical arrows represent the Ga surface diffusion regions. Consistent with (c) and (d), MBE experiments manifest the evolution of GaAs nanostructures (d) from quantum dots and quantum rings and double rings from 150 °C, 200 °C to 300 °C under the same As flux BEP of $1 \times 10^{-6}$ Torr with the same Ga amount of 6 ML; (e) from quantum dots to quantum rings by lowering the As flus BEP from $2 \times 10^{-4}$ Torr to $1 \times 10^{-5}$ Torr. (g) Schematic representation of the evolution from Frank-van der Merwe (layer by layer) mode to Stranski-Krastanow mode by making an As rich substrate surface.

Besides the droplet epitaxy of 0-D compound quantum nanostructures, metal droplets are also capable of yielding IV group (Si and Ge) nanodots or dendrites, in manner of metal induced crystallization (MIC).[103-107] A major feature of such droplet-based MIC process is that the precursors are provided in solid phase, either from the substrates as reservoirs, either by predeposition before the metal-semiconductor (M-S) interaction upon annealing. Au-Si interdiffusion has been studied for decades.[103, 104, 108, 109] *In situ* low energy electron microscopy (LEEM) study reveals the Au droplet reactive wetting on Si (111) surface, with Au atom surface diffusion and the formation of metastable Au silicide on the surface, as seen in Figure 3 (c).[110] Moreover, this reactive wetting behavior enables Si surface reconstruction into Si nanodots (see Figure 3 (b)), whose physical process is illustrated in Figure 3 (a).[111] On the other hand, once the metal droplets interacts with an amorphous coating layer atop the solid surfaces, dendritic crystals are often acquired, as seen in Figure 3 (d) of an Au-Ge case (the inset image shows the distribution of Au NPs), and in Figure 3 (e) of a Sn-Si case where Sn droplets can spread and be dissolved *en masse* by the amorphous Si layer (lower image).[112, 113]

One may be confused that such droplet-based MIC processes have little relevance with the droplet transport under vapor deposition. However, the solid precursors, especially of those predeposited amorphous coating layers, can indeed be regarded as a genre of infinitively fast vapor deposition on droplets, where the crystallization and the growth of low-dimensional materials induced by droplet reactive wetting mainly occur after the depositions. Moreover, it is also very interesting of the evolution from nanodots to dendrites when droplets wet substrate surfaces modified by amorphous layers. These issues will be discussed in the next section.

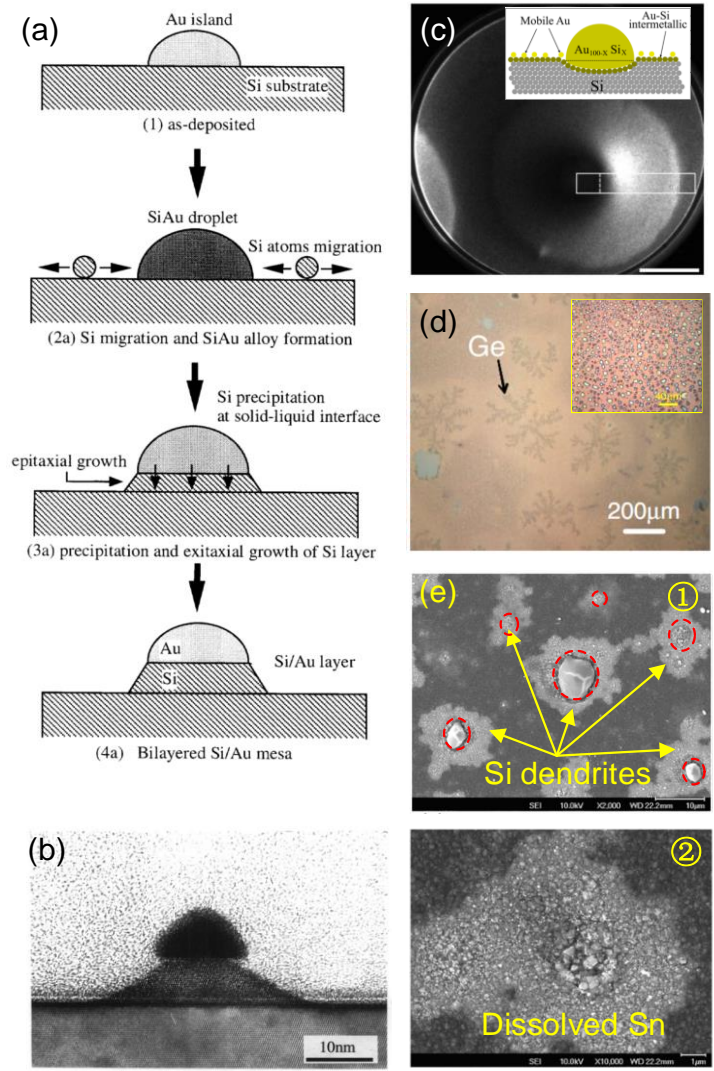

Figure 3. Metal droplet induced crystallization of IV group nanodots and dendrites. (a) Schematic representation of selective area growth of Si nanodots upon Au droplet epitaxy and (b) a corresponding cross-sectional TEM of bilayer of Au/Si nanodots. (c) a LEEM image of reactive wetting of Au on Si (111) surface, the inset image illustrates the wetting process, where Au droplet forms Au-Si eutectic alloy sits in thermodynamic equilibrium (the dark part), with a monolayer of metastable Au silicide surrounding it and with mobile Au atoms diffusing away from the bulk droplet (the light part in shape of annulus). (d) Optical image of crystalline Ge dendrite upon sputtering Ge on a heated Au thin film at 240 °C, which also turns Au into discontinuous islands (see the inset image). (e) (1) SEM image of crystalline Si dendrites growth around Sn islands upon heating Sn/Si bilayer, (2) probably with Sn droplets spreading and dissolved into dendrite areas *en masse*.

## 4. Droplet surface migration under vapor deposition for producing 1-D nanomaterials

The bottom-up approaches of producing 1-D nanomaterials date back to the VLS method for vertical Si whiskers growth proposed by Wanger et al.[114, 115] In such a growth process, the gaseous precursors keep feeding the metal catalytic droplets until supersaturation, which thereby activates the heterogeneous nucleation at the droplet-substrate interfaces. Very often, the droplets will be lift up from the substrates and lead the out-of-plane growth. Even though this vertical growth mode gives rise to the development of 3-D architectural devices, the planar ones are also in demand, for the sake of their diverse applications and of their van der Waals integration with other low-dimensional nanomaterials.[116-118] Thus, it is of particular importance to exploit the mechanism of 1-D nanomaterials growth via droplet in-plane transport.

### 4.1. In-plane solid-liquid-solid process

Prior to the discussion of the in-plane VLS process, we would firstly introduce a special growth process, that is, the in-plane solid-liquid-solid (SLS) growth process, which is applicable for Si, Ge or SiGe NWs growth. The first 'S' denotes the solid precursors (e.g. $a$-Si:H), which is predeposited before the NW growth. Let us, again, regard the hypothetical proximation that the predeposition on metal catalysts approaches a process of ultra-fast vapor deposition on droplets, which may favor a better understanding of the link between in-plane SLS and in-plane VLS in the following discussion.

Figure 4 (a) provides a typical in-plane SLS Si NW grown on $a$-Si:H coated $SiO_2$ substrate, where the In droplet migrates on the substrate by dissolving Si atoms from the $a$-Si:H thin layer and precipitate Si NW behind it via a continuous process of heterogeneous nucleation and crystal growth. We highlight the footprint left by the In droplet in-plane transport, called trench, as seen in the inset image of Figure 4 (a), whose cross-sectional profile is characterized by TEM-EDX mapping in Figure 4 (b). The roughly cylindrical Si NW is well located in the center of the trench, with residual $a$-Si:H at its both sides. Moreover, once Si (100) substrate instead of $SiO_2$ is applied, epitaxial Si NW will be obtained with a flat cross-sectional profile, as seen in Figure 4 (e) and (f). This must be attributed to the higher surface energy of Si (100) surface in comparison with the one of $SiO_2$, which gives rise to the wetting of Si nuclei on the Si (100) surface considering the system Gibbs free energy minimization.

In addition, such epitaxial NWs probably go straightforward, with no guiding treatment (see the inset image of Figure 4 (f-1)). Similar phenomena also occur on sapphire substrate, which will be discussed later. The in-plane SLS growth mechanism arises from the break of system thermodynamic equilibrium, as illustrated in Figure (c), where the Gibbs free energy difference between *a*-Si:H and *c*-Si, as the strains and distorted bonds in the amorphous matrix of Si provides excess free energy. A direct evidence is shown in Figure 4 (d), where no NW is produced once *μc*-Si:H instead of *a*-Si:H is deposited.

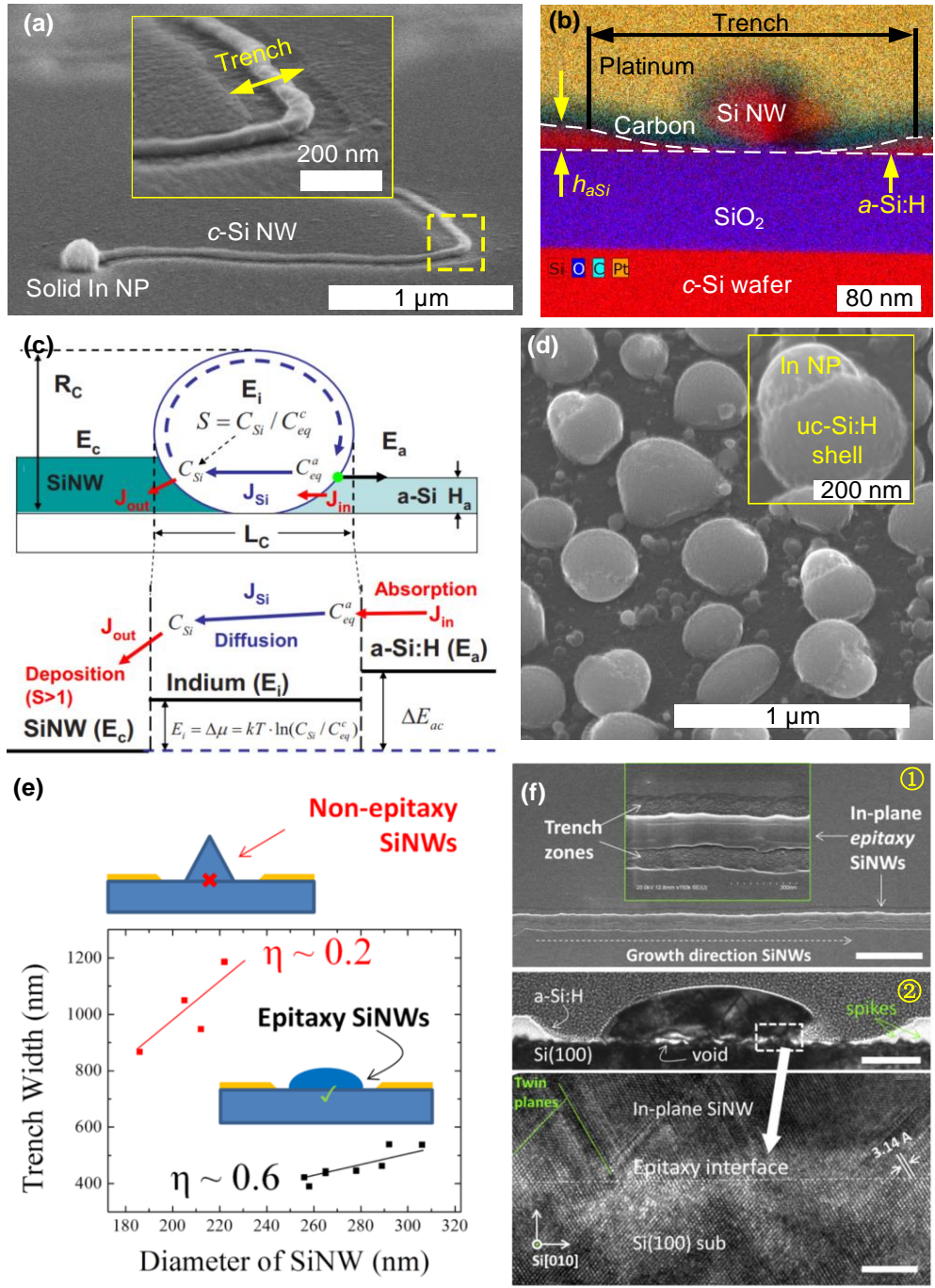

Figure 4. In-plane solid-liquid-solid growth mode of Si NWs. (a) SEM image of a typical randomly grown in-plane SLS Si NW on *a*-Si:H coated SiO$_2$ substrate, located in the trench which represents the footprint of the self-migrating In droplet (inset image). (b) Cross-sectional EDX mapping of an in-plane SLS Si NW on SiO$_2$, with highlighting the nearly cylindrical profile of the NW and the trench profile of *a*-Si:H. (c) Schematic representation that depicts the Gibbs free energy difference between *a*-Si:H and *c*-Si NW as the thermodynamic driving force of the in-plane SLS Si NWs growth. (d) SEM

image of the failure of in-plane SLS Si NWs growth by depositing *μc*-Si:H instead of *a*-Si:H, which is in consistent with (c). (e) Comparison of the ratios of NW diameter over trench width in non-epitaxial Si NWs grown on SiO$_2$ substrates and epitaxial Si NWs grown on Si (100) substrates, implying (f) flat NWs can be acquired on Si (100) surface with higher surface energy.

However, questions may arise that such phase transitions driven by the unbalanced thermodynamics generally occur in metal-semiconductor contact, as discussed in Section 2 (see Figure 3). The formation of either a nanodot, a dendrite or a NW, is supposed to be governed by the droplet in-plane transport behaviors, where *in situ* transmission electron microscope (TEM) shall be a powerful tool for the exploration.[119-138] The direction observation of in-plane SLS Si NW growth is recently realized via *in situ* TEM. Figure 5 (a) shows an as-grown NW on 40 nm thick *a*-Si:H coated amorphous SiN$_x$ membrane with an In NP in size of ~ 200 nm at its end, from which the *in situ* observation starts. The selective area electron diffraction pattern shows a [110] growth orientation at the NW front end. Figure 5 (b) and (c) display two sequences of TEM images extracted from *in situ* TEM video of at 350 °C, which records a rich interplay between the In droplet, the *a*-Si:H coated membrane and the Si NW. From $t$=2 s, the droplet starts to melt and alloy with the *a*-Si:H coating layer via interdiffusion. The initiation of the droplet motion is somehow dampened by the two arms, as seen at $t$= 4 s. However, its wetting layer starts to appear, which firstly propagates laterally from its two wings ($t$=4 to 6 s), then evolves into a hemi-annulus surrounding the droplet at $t$=8 s. During this period, new phase of *c*-Si is produced (colored in red). From $t$= 10 to 14 s, the droplet-NW interface clearly develops into three facets, which leads the *c*-Si precipitation in three directions with broad and polygonal NW growth (see the yellow part). Interestingly, once the droplet manages to escape two arms, it rapidly relaxes driven by its surface tension (see $t$=10 s and 14 s), with the trench borders exposed (marked by the black dashed lines). However, from $t$= 15 to 19 s, the droplet mainly propagates along the facets 1 and 2, while the facet 3 stops growing, resulting in a turning to right and even truncated NW morphology (see the green part) and the remarkably elongated droplet. From $t$= 19 s, the NW growth along facet 3 restarts, with a further stretched droplet that finally bursts for the first time at $t$= 21 s. This phenomenon of droplet breaking up into a bulk with tiny satellite dots can be regarded as a result of the competition between its surface tension and its continuous elongation induced by reactive wetting, which commonly appears in low-Reynolds-number droplet nanofluids.[139-142] Afterwards, the droplet relaxes and accumulates to its left side at $t$=22 s, where another self-turning proceeds. A more exaggerated droplet deformation with the formation of $c$-Si flake is recorded from $t$= 50 to 72 s, initiated by the same NW truncating and terminated by droplet breaking and turning. It is worth mentioning that the second TEM sequence in Figure 5 (c) indeed rules out the pinning effect of the arms on the droplet elongation and the consequent evolution from truncated NW to flake (see $t$=4 s in Figure 5 (b)), as there are no external forces applying on the droplet. The main factor should be the temperature that affects the reactive wetting induced droplet transport, as such phenomena disappear once the membrane is heated to 400 °C.

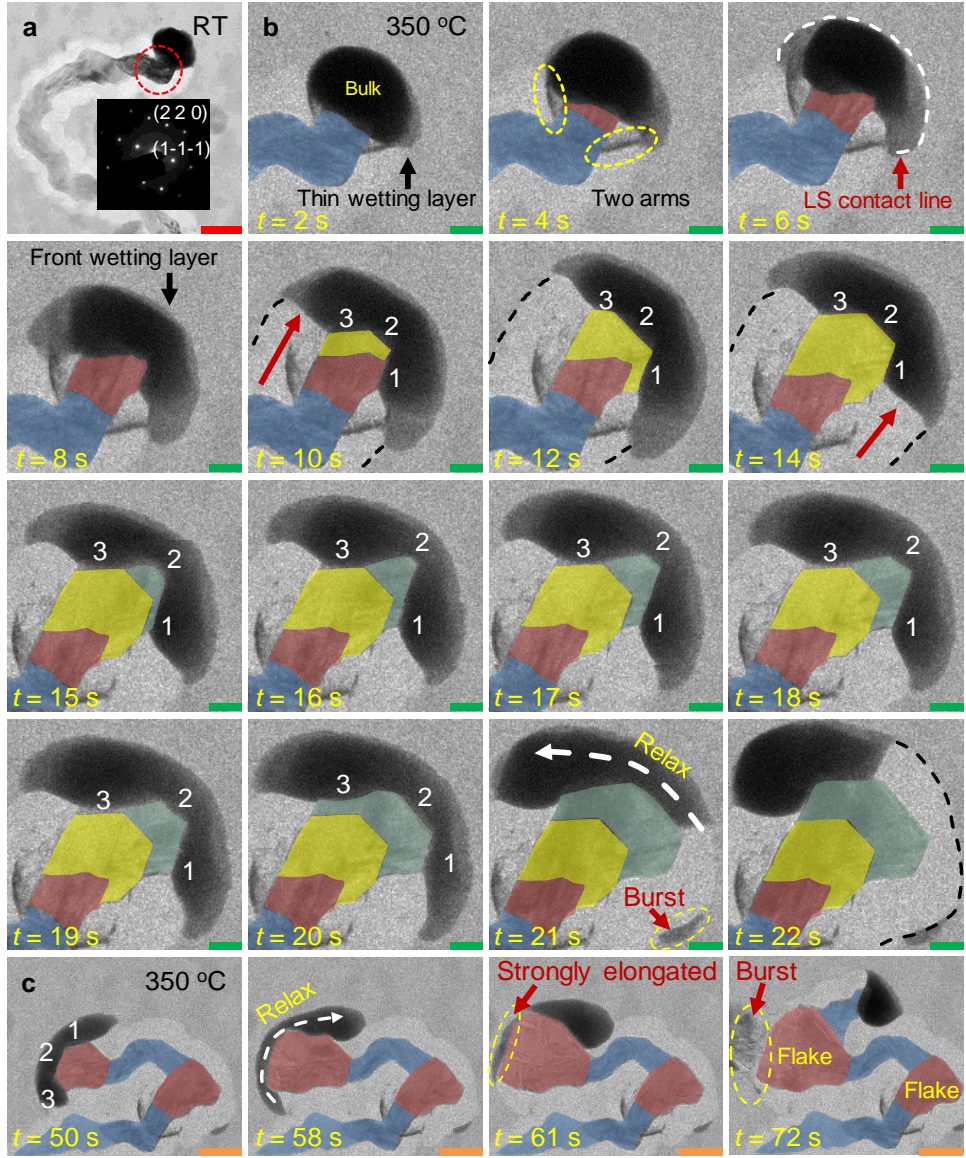

Figure 5. Two sequences of TEM images extracted from *in situ* TEM video of in-plane SLS Si NW growth at 350 °C. The scale bars are 200 nm, 100 nm, 400 nm in (a), (b) and (c), respectively

Before the *in situ* observation at 400 °C, the membrane temperature is held at 300 °C. Surprisingly, the droplet is stationary on the membrane with a wetting layer along the frontier S-L contact line, while no wetting behavior and *c*-Si precipitation are observed at its back, as seen at *t*=0 s in Figure 6 (a). This directly unveils the ***indium-philic*** properties of the *a*-Si:H coated surface, in comparison with the ***indium-phobic*** membrane surface deposited of *c*-Si precipitates, i.e.

$$\gamma_{aSi/sub} > \gamma_{cSi/sub}, (8)$$

where $\gamma_{aSi/sub}$ and $\gamma_{cSi/sub}$ denote the surface energy of a substrate (e.g. $SiO_2$, $SiN_x$, etc. that do not react with In droplets) whose surface is modified by deposited *a*-Si:H and precipitated *c*-Si, respectively. Note that *a*Si/sub and *c*Si/sub are taken into account as individual systems as a thin film or a NW cannot be considered as bulk materials. On the other hand, reactions (i.e. *c*-Si crystallization in this case) will cause the loss of system Gibbs free energy, which is mainly contributed by the solid surface energy decrease (see the reactive wetting part in Section 2).[64] In addition, the metastable wetting layer at the front is probably due to the insufficient amount of Si atoms dissolved within the droplet to approach the critical supersaturation for the nucleation activation.[143]

Once heating the membrane to 400 °C, the droplet spreads into a fan-like shape and its in-plane transport kicks off (see Figure 6 (b)). In contrast to the multi-faceting NW growth by the elongated droplet at 350 °C, the geometries of the droplet and of the droplet-NW interface are considerably maintained, which therefore produces a relatively uniform NW. On the other hand, random turning during the droplet surface migration is not eliminated, as seen from *t*=19 s to 23 s. This is very similar with the aqueous droplet free running reported by Dos Santos et al., which is associated with chemical inhomogeneity on the substrate surface, that is, the droplet spreads to the more wettable surface.[46] An alternative case of droplet turning (see Figure 6 (c)) occurs once it is trapped in a situation where *a*-Si:H is only present to its right side (see *t* =45 s, the black dashed line as the trench boundaries), then it spreads and turns towards the *a*-Si:H covered surface (see *t*=46, 47 s). This self-avoiding of its moving trace is also a remarkable signature of the droplet in-plane transport induced by progressive reactive wetting.[44]

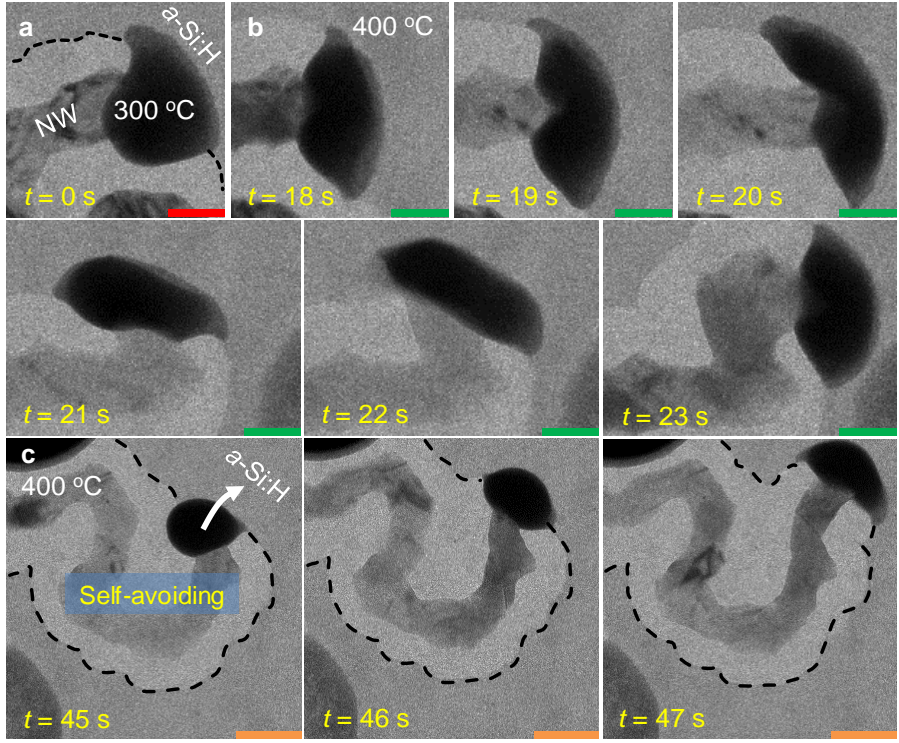

Figure 6. A sequence of TEM images extracted from *in situ* TEM video of in-plane SLS Si NW growth at 400 °C. The scale bars are 150 nm in (a), (b) and 500 nm in (c), respectively.

Apart from the *in situ* TEM observation, the geometric studies of the in-plane SLS system provides additional information about the droplet reactive wetting behavior. A top view scheme is illustrated in Figure 7 (a), with several parameters marked for the measurement: $d_{NP}$ denoting the solid In NP diameter as a sphere [144], $d_{In\text{-}aSi}$ the trench width, $d_{NW}$ the NW diameter, and $h_{aSi}$ the *a*-Si:H thickness (see Figure 4 (b)). From the viewpoint of droplet surface migration on solid surface, let us regard the NW as the byproduct of the droplet in-plane transport, like those discussed in the Section 2 (see Figure 1 (g), (f)), thus the most interesting parameters shall be the ones that determine the interaction between the droplet and the *a*-Si:H, i.e. $d_{NP}$ that relates to the droplet size), $h_{aSi}$ that gives rise to the reactive wetting, and $d_{In\text{-}aSi}$ as the footprint of the moving droplet. Figure 7 (b) provides the relationship between the trench width and the *a*-Si:H thickness (SiO$_2$ as substrate), after the normalization by the corresponding droplet size. The linear fitting is expressed as

$$\beta = 1.26 + 3.2\alpha, \quad (9)$$

where $\alpha$ denotes $h_{aSi}/d_{NP}$ that reflects the relative thickness of the *a*-Si:H with respect to droplet size,

and $\beta$ denotes $d_{In-aSi}/d_{NP}$ that reflects the degree of the droplet elongation attributed by the reactive wetting. This directly tells that, on the one hand, for the In droplet on bare $SiO_2$ surface, i.e. $\alpha=0$, the droplet-$SiO_2$ contact width (i.e. two times of the contact radius) equals 1.26, which means a hemi-spherical cap. This is in a nice agreement with the *in situ* SEM and TEM observation shown in Figure 7 (d).[145, 146] On the other hand, being a monotonically increasing relationship, the relatively thicker the *a*-Si:H is, the stronger the stretching is. Figure 7 (c) depicts the evolution of an In NP from a solid sphere, to a hemi-spherical droplet on bare $SiO_2$ surface and an elongated droplet arisen by its reactive wetting on *a*-Si:H coated $SiO_2$ surface. This could clearly explain the coalescence of In droplets during the ALC (amorphous-liquid-crystallization).[147] Similarly, in an in-plane SLS process, as seen in Figure 7 (e), the small In NPs within the 5 nm thick pad (see the right inset image) finally agglomerate into much larger ones by reactive wetting induced elongation and mutual contact, which thereafter lead the growth of NWs out of the In pad zone. Note that the sample is treated by a $SF_6/CHF_3$ plasma etching of Si, which helps unveiling the droplet coalescence phenomenon, the traces of the Si NWs are marked by the yellow dashed line in the left inset image.

Based on the *in situ* TEM observation and geometric studies, a general picture of Si NW growth by In droplet in-plane transport on *a*-Si:H coated substrate surface is drawn in Figure 7 (g), by compiling the knowledges of reactive wetting and droplet self-propulsion (see Figure 1), that is, (i) the In droplet reactively wets the *a*-Si:H coated substrate surface with Si-In atomic interdiffusion, (ii) heterogeneous nucleation of *c*-Si on the substrate surface will take place once the droplet gets supersaturated, (iii) thereby the droplet starts to move due to the establishment of a wettability gradient from the substrate surface with *c*-Si precipitated to that coated by *a*-Si:H. All these three steps in cycles will enable a continuous droplet in-plane transport as well as 1-D NW growth. In addition, a Si concentration gradient should exist within the droplet, which drives the Si diffusion from its advancing side where Si atoms are continuously supplied, to its receding side where Si atoms are kept being consumed by nucleating at the TPL with the following ledge flow mechanism [97, 134, 136, 148]. However, we would clarify two issues here:

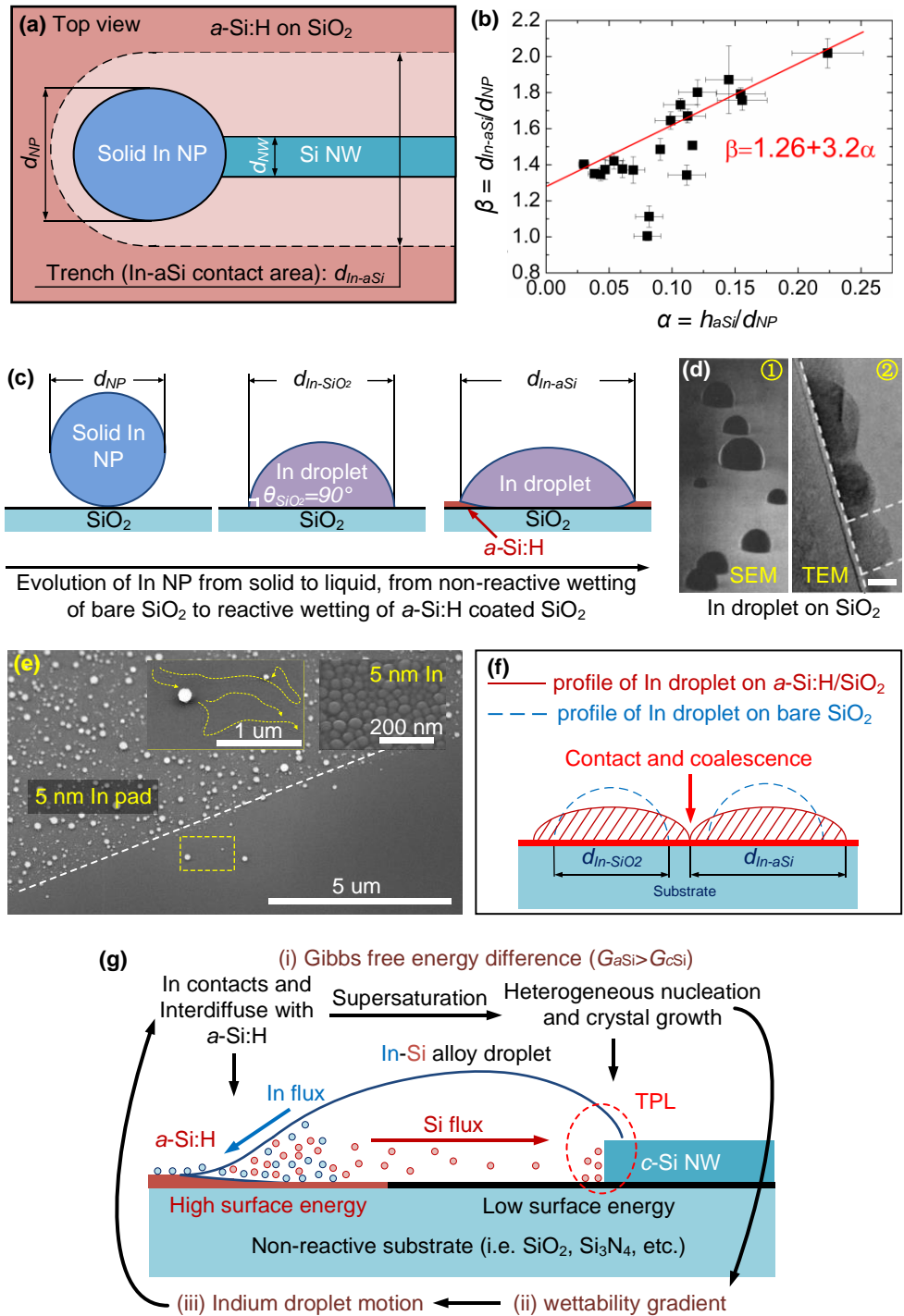

Figure 7. Reactive wetting induced In droplet in-plane transport for in-plane SLS Si NW growth. (a) Top view scheme of an in-plane SLS Si NW with the geometric parameters. (b) Plotting of the trench width as a function of *a*-Si:H thickness, with both parameters normalized with the corresponding In NP sizes. The linear fitting result reveals the reactive wetting behavior of In droplets on *a*-Si:H coated substrate surface, which does not appear for the ones on bare $SiO_2$ surface. (c) Schematic

representation of the evolution of In droplet from a spherical solid to hemispherical droplet on bare $SiO_2$ and elongated spherical cap on *a*-Si:H coated $SiO_2$. (d) SEM and TEM images of In droplet with ~90° contact angle. (e) Size redistribution of In droplets at the initial stage of in-plane SLS Si NWs, which results in larger droplets that lead the NW growth, which is caused by the reactive wetting induced contact-and-coalescence process, as illustrated in (f). (g) Schematic representation of the in-plane SLS growth mechanism.

However, such a picture does not cover all the aspects, there remain two interesting questions relevant to the droplet wetting: why is the droplet elongation stronger (1) at relatively lower temperature (350 °C) which can be completely eliminated at 400 °C, (2) or on the relatively thicker *a*-Si:H coated surface? Moreover, strongly stretched droplet will form multi-interfaces with the NW, which makes the NW truncating (even evolving into flakes) and induces the droplet swing during its in-plane transport. Recent studies on the in-plane SLS process by other metal/semiconductor pairs may provide clues. Figure 8 display the in-plane SLS grown (a) Ge NWs by Sn droplets[149], (b, c) Si NWs by Sn droplets[150], and (d) Ge NWs by In droplets, respectively. Table 1 summarized the growth parameters of all the in-plane NWs discussed above. Apart from the differences of the amorphous layer thicknesses and the droplet sizes, a common feature of the three is that the serpentine NWs have tails with droplets totally consumed, which is the most significant difference with the Si NWs mediated by the In droplets (see Figure 4 (a)). In particular, many tiny Sn dots are clearly seen in Figure 8 (a-1), implying that the mass loss of the droplets is probably attributed by their break-up during in-plane transport, similar with the phenomenon *in situ* observed at 350 °C (see Figure 5). Due to the dissipation of droplets during their motion, they cannot migrate long distance for long NWs (e.g. the Ge NWs by Sn or In as seen Figure 8 (a) and (d)), or they need large size (in diameters of 200~600 nm) to compensate the mass loss during the NWs growth (e.g. Si NWs by Sn as seen in Figure 8 (b, c)). It has been mentioned in Figure 5 that moving droplets are easy to break due to the competition between their surface tensions and deformations.[142] However, In and Sn are very close to each other on properties like surface tension, viscosity and density. Thus, there should be other factors that prevails for determining the different droplet in-plane transports of the Sn-Si, the Sn-Ge and the In-Ge pairs with respect to the In-Si pair. The solubilities of S (for semiconductor) in M (for liquid metal droplet) may be one factor. Figure 8 (e) summarizes several $C_{s-m}$ (i.e. atomic concentration percentage at.%) at their individual eutectic points and in the range of the growth temperature (300 °C ~ 400 °C) in equilibrium phases.[151-156] [157] Note that the values of $C_{Si-In}$, $C_{Si-Sn}$, $C_{Si-Au}$ and $C_{Si-Pb}$ in the range of growth temperature are not precisely provided in literatures, but they vary very little in the vicinity of their eutectic points. One can see that Si has the lowest concentration in liquid In ($5\times10^{-10}$ ~ $1\times10^{-9}$ at.%), which is several orders of magnitudes lower than those of Si in Au, Si in Sn, Ge in Sn and Ge in In. Being a process of nucleation from solution, the difference between semiconductor chemical potential $\Delta\mu$ in supersaturated and saturated metal droplets drives the nucleation and NW growth, which is expressed as[158]

$$\Delta\mu = k_B T ln\left(\frac{C_s}{C_0}\right) = k_B T ln\beta, (10)$$

where $k_B$ denotes the Boltzmann constant, $T$ the Kelvin temperature, $C_s$ and $C_0$ the semiconductor concentration in supersaturated and saturated metal droplets, respectively, and the supersaturation degree can be represented by $\beta=C_s/C_0$. This implies that, considering the same semiconductor concentrations in the supersaturated metal droplets, the Si-In pair will have the highest $\beta_{Si-In}$ as well as $\Delta\mu_{(Si-In)}$, thereby render the lowest critical Gibbs free energy for nucleation $\Delta G^*_{heter(Si-In)}$ and the fastest nucleation rate $J_{Si-In}$, whose relationships are given as[143, 158]

$$\left.\begin{aligned}\Delta G^*_{homo} &= \frac{16\pi\Omega^3\gamma^3}{3(\Delta\mu)^2}\\ \Delta G^*_{heter} &= \Delta G^*_{homo}(\frac{1}{2}-\frac{3}{4}cos\theta+\frac{1}{4}cos^3\theta)\\ J &= J_0 exp(-\frac{\Delta G^*}{k_B T})\end{aligned}\right\}, (11)$$

where $\Delta G^*_{homo}$ and $\Delta G^*_{heter}$ denote the critical Gibbs free energy for homogeneous and heterogeneous nucleation, respectively; $\gamma$ the surface energy of the nuclei, $\Omega$ the volume of a semiconductor atom, $\theta$ the contact angle of nuclei on substrate surface, $J$ the nucleation rate, $J_0$ the frequency factor. Note that this set of calculations is by considering the nucleus as a sphere for homogeneous nucleation and a spherical cap for heterogeneous nucleation. By translating Equation (11), the heterogeneous nucleation rate $J_{heter}$ can be written by

$$J_{heter} = J_0 \exp\left[-\frac{16\pi}{3} \cdot \left(\frac{\Omega\gamma}{k_B T}\right)^3 \cdot (\ln\beta)^{-2} \cdot g(\alpha)\right] \Biggr\}, (12)$$
$$g(\alpha) = \frac{1}{2} - \frac{3}{4}\cos\theta + \frac{1}{4}\cos^3\theta$$

where $g(\theta)$ is the geometric factor in function of the contact angle $\alpha$ between the nucleus and the substrate surface.

In association with the RPC model of reactive wetting (see Section 2, Figure 1 (a)), the droplet in-plane transport for NW growth is supposed to be nucleation limited[159], in contrast with a VLS process that is mainly governed by the precursor adsorption/diffusion limited model.[97, 98, 134, 136, 148, 160-164] Thus, a fast nucleation favors a fast establishment of wettability gradient on substrate surface, thereby enables the droplet making steps forward fast in each cycle of the stick-slip motion[44]. By associating Equation (12), one can have

$$v_{dp} = v_{\nabla\gamma} = r_{sls} \sim J_{heter}, (13)$$

where $v_{dp}$ denotes the droplet velocity, $v_{\nabla\gamma}$ the speed of the surface energy gradient establishment for each step of the crawling droplet, $r_{sls}$ the growth rate of an in-plane SLS NW. Otherwise, for other S-M pairs with high $C_0$, the droplets have to wet the amorphous layer more strongly in attempt to collect enough semiconductor atoms (i.e. larger $C_s$) for achieving the critical supersaturation $\beta$, which brings about risks of burst during their motion.

Likewise, according to Equation (10), once the temperature $T$ is lowered, a higher supersaturation $\beta$ should be acquired for the compensation of the free energy loss in the M-S systems. This may account for the strong spreading of In droplet at 350 °C (see Figure 5), in the same reason of contacting and dissolving more Si atoms. On the other hand, regarding the geometric factor $g(\theta)$ of $\Delta G^*_{heter}$ in Equations (11, 12), $g(\theta)$ is monotonically decreasing as a function of contact angle in range of $[0, \pi]$, thus the formation of Si flake is more favorable at low temperatures, as it is easier for a flake with low contact angle ($\theta\rightarrow 0$) to overcome the nucleation barrier (i.e. $\Delta G^*_{heter}$) than a NW ($\theta\rightarrow\pi$).

Last, given a droplet in a constant size, if the amorphous layer turns thicker, the droplet has to spread and swallow more semiconductor atoms to deplete the amorphous layer and in consequence make the substrate surface exposed for the following heterogeneous nucleation (see Figure 4 (b) and (f)). Thus, the stretching effect on the droplet shall be enhanced, as seen in Figure 7 (b).

Based on the above discussion, one can see that the droplet reactive wetting will be enhanced in conditions of low $T$, higher $C_0$ or higher $h_{aSi}/d_{NP}$, which makes the droplet vulnerable to breakup during its in-plane transport for in-plane SLS NWs growth. Moreover, each of the three factors, or in certain combinations, will lead a multi-faceted droplet-NW interface along which a frequent swing of the droplet may occur. More interestingly, once a high index facet is formed, the droplet will wet it due to its higher surface energy[165-167], accompanied by a crystallographic-index-lowering self-turning sequence for zigzag NW growth, as seen in Figure 8 (c).[150] In addition, Figure 8 (e) provides the information of Si-Ga and Si-Pd pairs, where the former one suggests that Ga is supposed to be an alternative candidate for in-plane Si NWs growth at low temperature due to the low Si solubility, while the latter one will be discussed in 4.2 in-plane VLS process.

**Table 1 Summary of growth parameters of in-plane Si or Ge NWs**

| No. | NW | NP catalysit | $h_{aSi}$ ($h_{aGe}$) /nm | Temperature | Note | Ref |
|---|---|---|---|---|---|---|
| #1 | Si (seldom turning) | In, ~ 100 nm (NW end) coalescing from ~ 30 nm In NPs in 5 nm thick In pad | 10 nm | 320 °C | Figure 4 (a), in PECVD | |
| #2 | Si (frequent turning, even into flake) | In, ~200 nm (NW end) | 40 nm | 350 °C | Figure 5, in TEM chamber | |
| #3 | Si (frequent turning) | In, ~200 nm (NW end) | 40 nm | 400 °C | Figure 6, in TEM chamber | |
| #4 | Si (periodic turning) | Sn, 200~600 nm (starting point) | 30 nm | 340 °C | Figure 8 (b,c), in PECVD | [150] |
| #5 | Ge (periodic turning) | Sn, up to 410 nm (starting point) | 14 nm | 270 °C | Figure 8 (a), in PECVD#1 | [149] |
| #6 | Ge | In, coalescing | 10 nm | 250 °C | Figure 8 (d), in PECVD | |

| No. | NW | NP catalysit | $h_{aSi}$ ($h_{aGe}$) /nm | Temperature | Note | Ref |
|---|---|---|---|---|---|---|
| | (periodic turning) | from ~ 30 nm In NPs in 5 nm thick In pad | | | | |

Note that the substrate temperatures of the PECVD systems are calibrated by thermal couple, the TEM heating membrane temperature is calibrated with In melting point as reference.

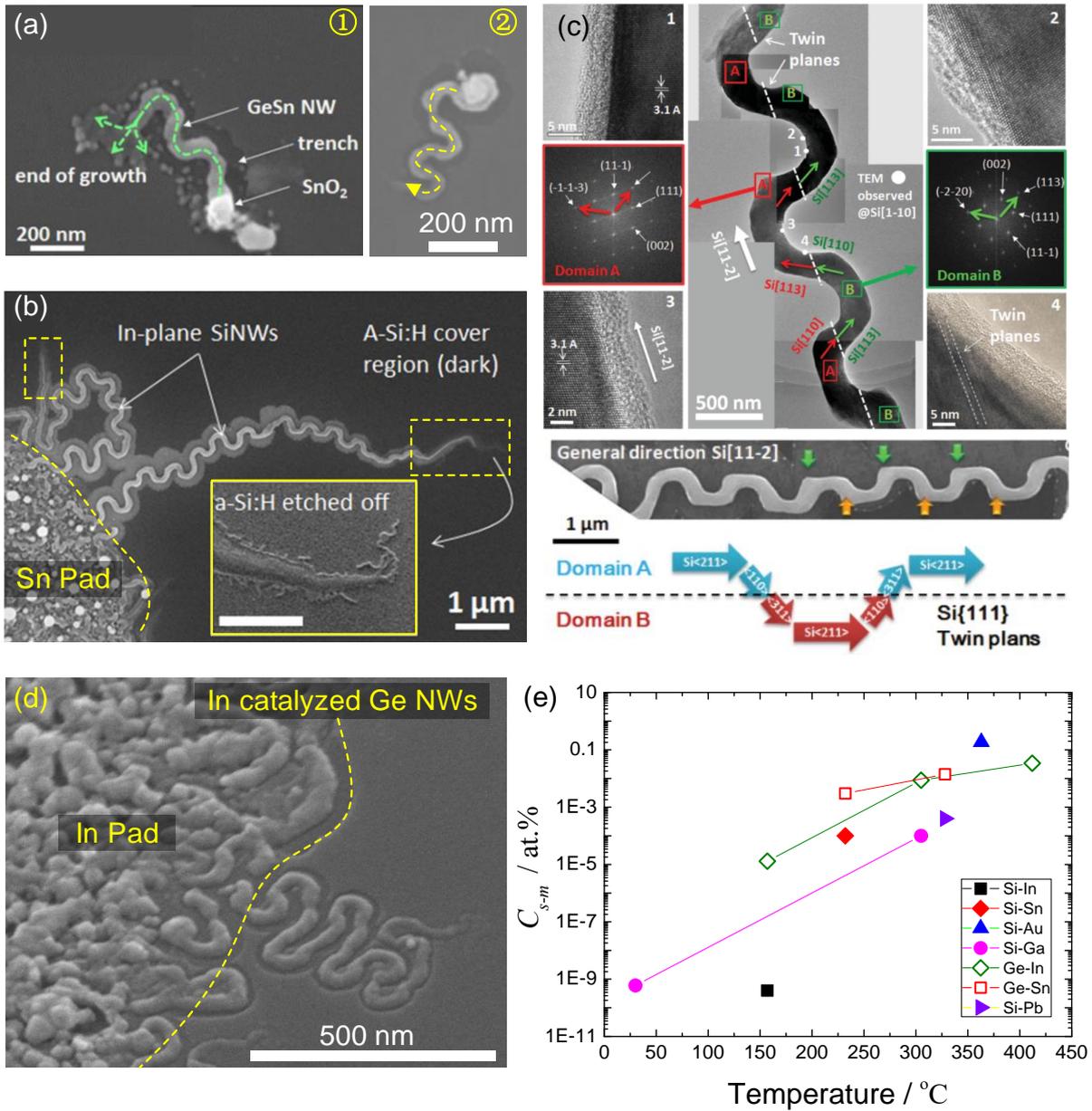

Figure 8. Zigzag NWs growth via in-plane SLS process. (a) Sn droplet catalyzed zigzag Ge NWs. (2)

Sn droplet catalyzed zigzag Si NWs, (c) which follows a crystallographic-index-lowing self-turning sequence revealed by TEM studies. (d) In droplet catalyzed zigzag Ge NW. (e) Summary of solubility of Si or Ge in different metals at their eutectic points.

4.2. In-plane vapor-liquid-solid process on planar substrates

In contrast to the in-plane SLS process for the planar NWs growth, the VLS process is mainly applied to produce out-of-plane ones, which changes little since its invention.[114, 168] A plausible interpretation would be that the SLS process provides solid precursors on the substrate prior to the NW growth which enables a 100% yield of in-plane NWs, whereas gaseous precursors are deposited *in situ* during the VLS growth process which gives rise of the droplet detachment from the substrate and maintains its out-of-plane transport. However, theoretical simulations predict that the in-plane transport of droplets under VLS growth conditions is possible, by intentionally perturbating the droplet-NW interfaces.[169-172] As seen in Figure 9 (a), (1) a vertical NW is produced via a default VLS process; however, (2) once an excess free energy is introduced on the right side of the NW sidewall surface, the droplet will spread to the higher surface energy surface and the NW kinking takes place; moreover, (3) at the initial stage of NW growth, if tuning the sidewall of the NW pedestal to be with high surface energy, the droplet will spread to the pedestal edge and the in-plane NW growth starts to proceed.[170]

The idea of the NW surface energy engineering is straight forward, while optimizing the VLS deposition conditions are tricky. We shall firstly discuss the in-plane VLS growth process on planar substrates without nanostructuration. Based on the in-plane SLS growth mechanism in Section 4.1, it is clear that In droplets reactively wet the *a*-Si:H coated surface. Thus, a facile method of growing in-plane VLS Si NWs is shown in Figure 9 (b), that is, coating a thin *a*-Si:H layer (e.g. in thickness of 10 nm) on In NPs below the In melting point (e.g. at 150 °C) before starting a normal VLS process at 400 °C. As seen in the SEM image in Figure 9 (b), in-plane SLS Si NWs are firstly grown during the substrate temperature ramping to 400 °C, then the VLS growth process starts and the VLS Si NWs incline to grow horizontally, with the droplets attaching the substrate surface (see the inset image) as a direct evidence. Besides ameliorating the wettability of substrate surface by precoating amorphous thin film, a single VLS process can also yield in-plane NWs. As seen in Figure 9 (c), under the same deposition conditions, Pb droplets moves on the substrate surface with Si NWs growth, while In droplets are lifted up with out-of-plane NWs grown.[173] This is probably because the solubility of Si in Pb is much higher than that in In (see Figure 8 (e)), which will retard the Si supersaturation and nucleation within the Pb droplets. During this incubation period, a thin film of *a*-Si:H has been deposited on the substrate surface, thereby pinning the droplets and inducing their in-plane transport. Similar phenomena occur for GeSn NWs, as seen in Figure 9 (d), where large Sn droplets with longer incubation time lead the growth of in-plane NWs while the small ones leave the substrate for out-of-plane transport.[174] However, controversies will arise since one may argue that Au was the first catalyst applied for out-of-plane VLS growth, which can accommodate much larger amount of Si atoms than Pb. We would mention that the vapor deposition rate is a key factor, as it is the competition between the growth of the NW pedestal growth and the thin film on the substrate that determines the droplet transport behavior during the NW growth, as illustrated in Figure 10. For the in-plane growth mode, we propose that before the detachment of the droplet by the growing NW pedestal, a continuous thin film with high wettability should cover the substrate surface in vicinity of the droplet, which enables the attainment of the droplet on the substrate surface for the subsequent in-plane transport. Thus, this criterion of the in-plane NW growth can be expressed by

$$\left.\begin{matrix} r_{ped} < r_{flm} \\ \gamma_{ped} < \gamma_{flm} \end{matrix}\right\}, (14)$$

where $r_{ped}$ and $r_{flm}$ denote the growth rates of the NW pedestal and the thin film on the substrate surface, respectively; $\gamma_{ped}$ and $\gamma_{flm}$ denote the individual surface energies. Otherwise, once the droplet is lifted up from the substrate surface by the growing NW pedestal beneath it, i.e.

$$r_{ped} > r_{flm}, (15)$$

an out-of-plane growth will occur. Moreover, we propose that the in-plane SLS growth mode is a very particular type of the in-plane VLS growth mode, since it could be regarded as a deposition process of highly wettable thin film in an infinitively high rate and in an infinitively short period, during which the NW pedestal is hardly grown, which can be expressed as

$$\left.\begin{array}{r}r_{flm} \to +\infty \\ r_{ped} \to 0 \\ t_{dep} \to 0 \\ \gamma_{sub} < \gamma_{flm}\end{array}\right\}, (16)$$

where $\gamma_{sub}$ denotes the substrate surface energy and $t_{dep}$ the deposition time. Technically, this could be enabled by depositing the thin film prior to the NW growth (i.e. in-plane SLS process).

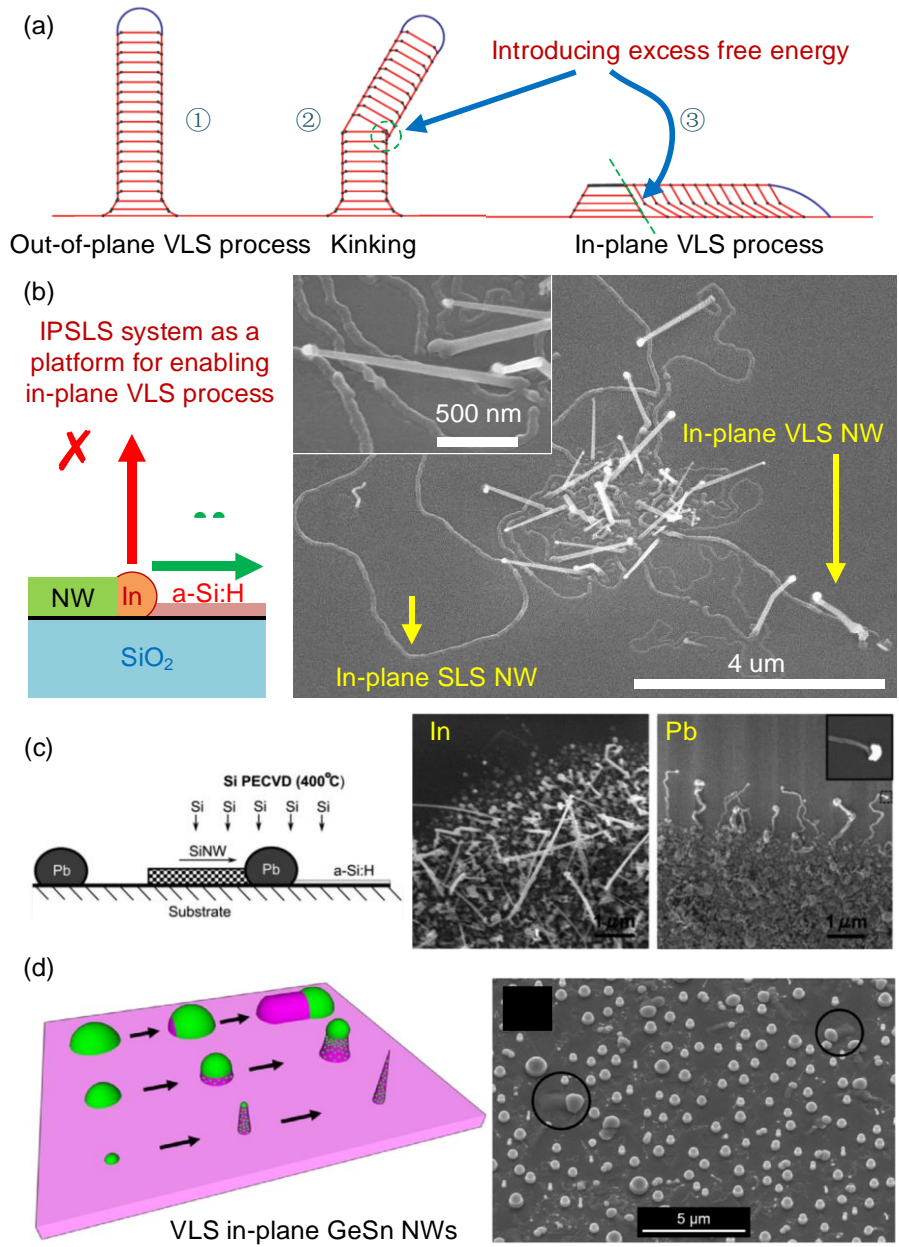

Figure 9. Droplet in-plane transport under VLS conditions for planar NWs growth. (a) Theoretical simulation of introducing excess droplet-NW interfacial energy during (1) a normal VLS process, which results in (2) NW kinking and (3) in-plane growth. (b) in-plane SLS system as a facile platform

that enables in-plane VLS process, as the *a*-Si:H coating layer turns the substrate more wettable and impedes the detachment of the droplet. (c) In-plane VLS growth of Si NWs by using Pb droplets as catalysts. (d) In-plane VLS growth of GeSn NWs by large Sn droplet, whereas the small ones follow the out-of-plane growth mode.

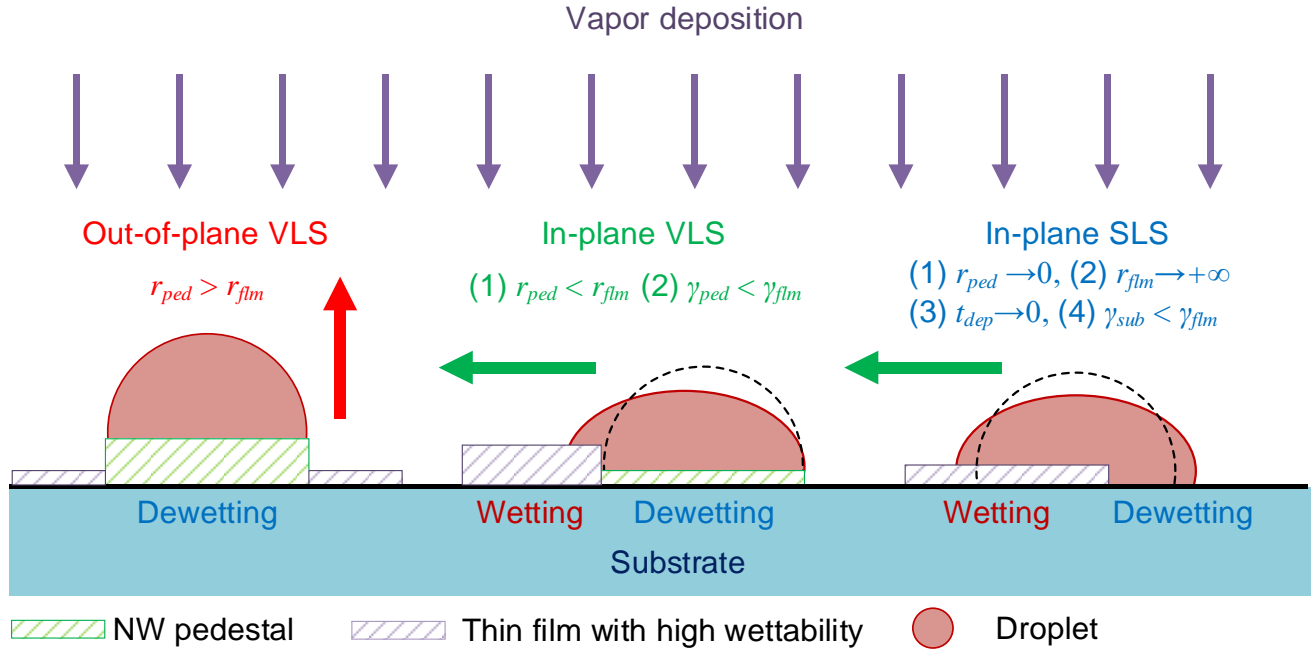

Figure 10. Schematic representation of catalytic droplet transport under vapor deposition for NWs growth. (a) If the formation of the NW pedestal is faster than the thin film deposition on the substrate (i.e. $r_{ped} > r_{flm}$), the droplet will be detached from the substrate surface and drive an out-of-plane NW growth. (b) If the thin film grows faster than the NW pedestal (i.e. $r_{ped} < r_{flm}$), meanwhile with a higher surface energy than the NW pedestal (i.e. $\gamma_{ped} < \gamma_{flm}$), the droplet will wet the thin film, which results in the in-plane NW growth. (c) If a thin film with a higher wettability compared with the substrate is deposited in an infinitively high rate and in an infinitively short period during which the NW pedestal is hardly grown (i.e. $r_{ped} \to 0$, $r_{flm} \to +\infty$, $t_{dep} \to 0$, $\gamma_{ped} < \gamma_{flm}$), which means the thin film deposition is prior to the NW growth, an in-plane SLS growth will take place.

4.3. In-plane guided transport of droplets on nanostructured substrates

In attempt to exploit more universal routes for achieving in-plane 1-D nanomaterials (e.g. NWs, CNTs) growth and their self-organization, the strategies of substrate nanostructuration are gaining a great interest. A review on this topic has been published elsewhere[116]. In this review, we would select several cases that involves interesting droplet in-plane transport behaviors.

4.3.1. Guiding nanochannel method

When the Au catalyst is buried in a nanochannel by the fabrication procedures illustrated in Figure 11 (a), the subsequent VLS process will produce a Si NW whose growth direction is guided by the nanochannel (see the SEM image in Figure 11 (b)).[175, 176] A direction impression is probably that the confined Au droplet has no choice but moves along the nanochannel and the inertia effect will keep its horizontal motion. However, we would mention that the $SiO_2$ surface with low surface energy is difficult to pin the droplet for an in-plane VLS process. Based on the conclusions from the in-plane SLS growth process in 4.1, a possible interpretation may be that before the droplet getting out of the nanochannel, the glass surface has been coated by an thin film of Si (probably *a*-Si:H) that is wettable by the droplet, otherwise the droplet has high opportunities to leave the substrate and proceeds an out-of-plane transport. This is very similar with the horizontal VLS grown Si NW on the *a*-Si:H coated substrate (see Figure 9 (b)). In addition, we estimate that the Si deposition rate should be be very high, otherwise the Au droplet will emerge into the thick Si layer with the formation of Si dendrites (see Figure 3 (d, e)).

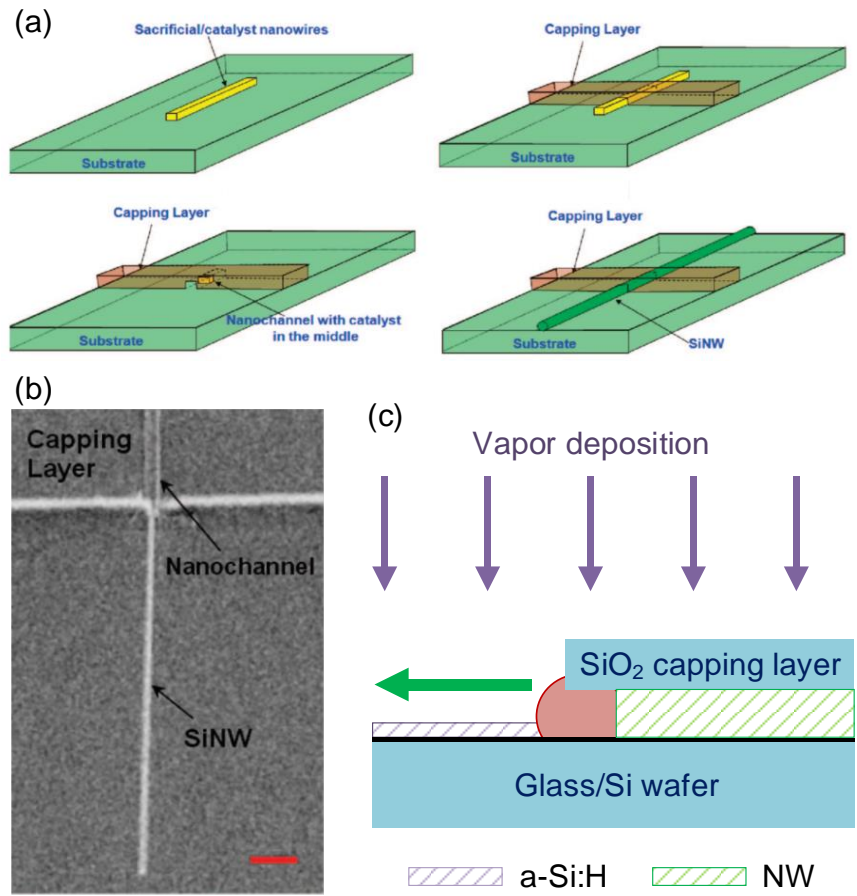

Figure 11. Nanochannel method for the guided VLS Si NW growth. (a) Fabrication flow chart of the guiding nanochannel on the substrate. (b) A SEM image of a typical guided VLS Si NW. (c) Schematic representation of a possible guided growth mechanism.

4.3.2. Substrate nanofaceting method

    The guiding nanochannel method is till now only demonstrated for the VLS growth of Si NWs. However, the nanofacets on the substrate surfaces are regarded as a more universal platform. The most prevailing strategy is called Graphoepitaxy by Joselevich et al., where different in-plane and self-aligned semiconductor NWs or CNTs have been realized on various types of substrates such as quartz, sapphire, SiC, spinel or Si.[177-186] Take Sapphire C-plane surface for instance, the nanofacets can be fabricated either by miscutting the substrate (Figure 12 (a-1)) or by high temperature annealing of well cutted substrate to make step flow (Figure 12 (a-3)), with the corresponding self-aligned GaN NWs as seen in Figure 12 (a-2, 4), repectively.[179] However, the out-of-plane growth process cannot

be completely eliminated. Drucker et al. studied the effect of the miscut angle and the deposition conditions on the in-plane growth yield. Figure 12 (b) show the VLS growth of Si NWs on Si (111) surface with 10° miscut angle: the large Au catalyst leads an in-plane in-plane NW growth while the small one goes out-of-plane for vertical NW growth. This is in good consistence with the Ge NWs catalyzed by Sn droplets (see Figure 9 (d)), as it takes less time for small droplets for the formation of NW pedestal so that the droplet can be detached from the substrate surface (see the scheme of the out-of-plane VLS process in Figure 10). The thermodynamic calculation indicates that the droplet can be either pinned at hilltop or valley bottom of the Si (111) nanofacets, where the S-L free energy reaches the minimum, as illustrated in Figure 12 (c). Moreover, statistical studies show that the yield of in-plane NWs can be improved on high miscut angle surface (i.e. 10° rather than 4°) and under low pressure disilane deposition, as seen in Figure 12 (d). For the former factor, since the Au droplet wets the well cutted Si (111) with a contact angle of 43° (i.e. acute angle)[187], so the Si (111) surface energy is higher than the L-S interfacial energy (see Equation (1) of the Young's equation). Thereby, the introduction of the nanofacets will bring about excess free energy, so the Au droplet has to wet more to compensate the increment of the system free energy. In other words, the wetting behavior of Au droplet is enhanced on miscutted Si surface. Let us consider the nanofacets as a kind of surface roughness, according to the studies of the effect of surface roughness on the contact angles, the higher the miscut angle is, the higher the wettability of the Si surface is.[188] Thus, the miscut angle of 10° should favor the droplet pinning and the consequential in-plane NW growth. For the latter factor, as illustrated in Figure 10, low disilane pressure delays the formation of NW pedestal which provides high chances for in-plane NW growth.

Other than the graphoepitaxy method whose crucial step is the substrate miscutting, the homoepitaxial growth of in-plane guided III-V NWs presents an alternative route without the pretreatment of the substrates.[189-191] We would mention that the technique seems different, while the guiding principle is indeed similar. A direct evidence is displayed in Figure 12 (g), one can see that the GaAs epilayers in morphologies of atomic steps or grooves are formed on the GaAs (110) substrates at 450 °C, by varying the V/III ratios and the epilayer growth rates: (1) V/III ratio of 3 and 0.1 μ/h, (2) V/III ratio of 20 and 0.1 μ/h, (3) V/III ratio of 3 and 0.5 μ/h, (4) V/III ratio of 3 and 1 μ/h,

(5) V/III ratio of 10 and 1 μ/h.[192] This result indicates that the substrate nanofacets are formed in a VS growth process in parallel with VLS growth of NWs. To suppress the out-of-plane growth, systematic studies of the V/III ratio have been carried out. Figure 12 (e) shows an array of self-aligned VLS grown GaAs on GaAs substrate with Au as catalyst. Two main factors are found that affect the yield of in-plane growth.[191] The plotting in Figure 12 (f) indicates that broad NW bottom width (>70 nm) favor the in-plane growth. This is in accordance with the Equation (14) as well as the results from Figure 9 (c, d), that is, the broad NW bottom width means a longer incubation time which allows the substrate surface modification by the deposited thin film (i.e. the growth of GaAs epilayer here) and thus provides high opportunity to pin the droplet by wetting. Unlike the reactive wettability of the amorphous Si (or Ge) thin films by the catalytic droplets, engineering the III-V surface energy relies on the fine tuning of the III/V ratio. The plotting in Figure 12 (f) displays the evolution of the yield of the planar GaAs NWs in relationship with the V/III ratio. When the As flux is very low (see the violet curve for the V/III ratio of 0.4), we suggest that the VLS growth surpass the VS growth (i.e. $r_{vls} > r_{vs}$ in Figure 10), thereby very probably the Au droplet has been detached from the substrate before the formation of the critical nanofacets for droplet pinning, since the discontinuous 2D islands are formed at the initial stage of the homoepitaxy before they coalesces and proceed the step flow process.[193, 194] Once the V/III ratio is raised up to 1.3, a noticeable improvement is achieved, with 100% success of the planar growth for NW bottom width above 70 nm (see the red curve). However, if the V flux is aggressively injected (i.e. the V/III ratio of 18), both of the epilayer and the NW will grow much faster, in particular for those in small sizes, which brings about uncertainties of the droplet pinning by the strained epilayer, thus the yield of in-plane growth drops with the decreasing NW bottom width (see the green curve).

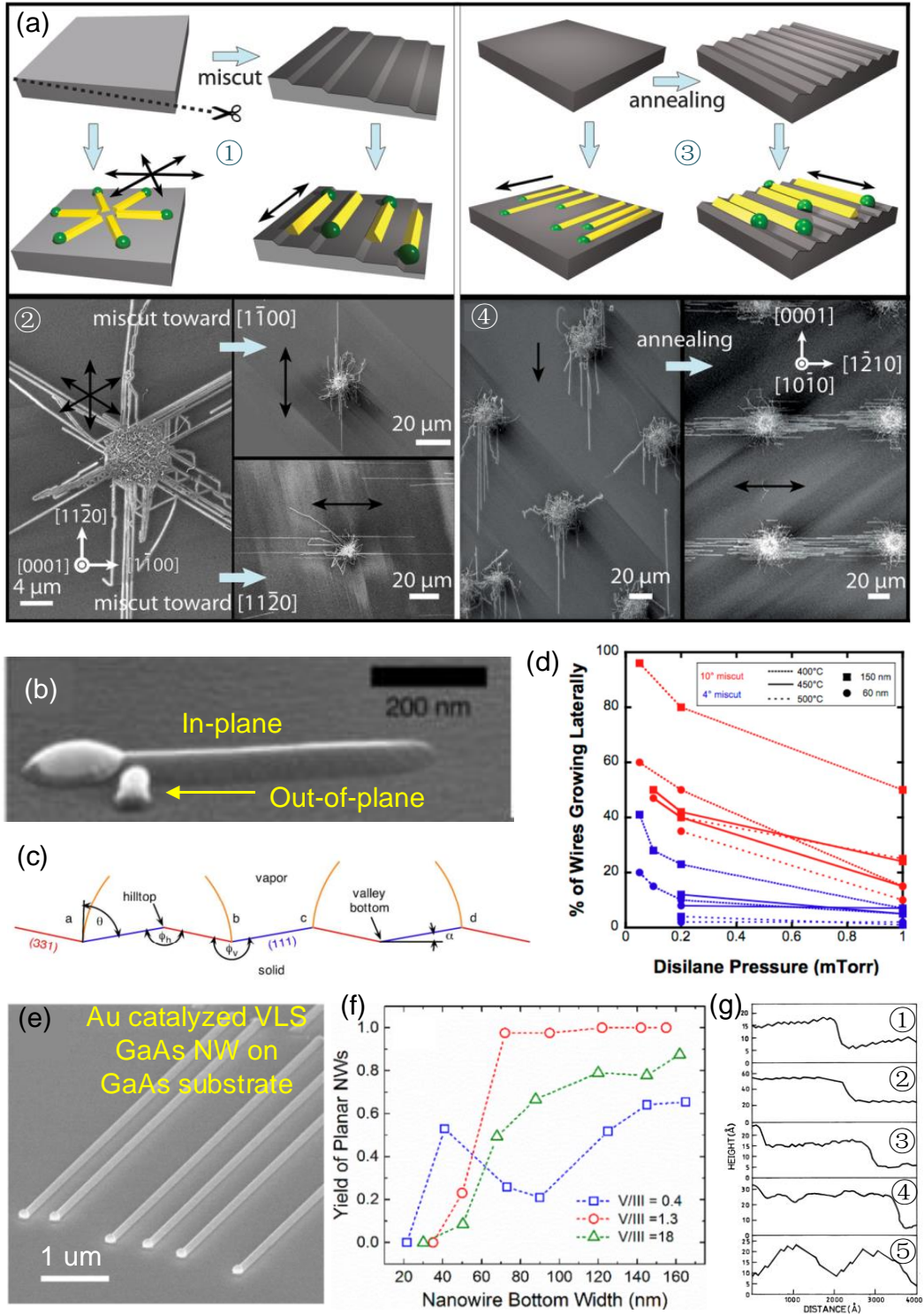

Figure 12. Graphoepitaxial method for guided VLS NWs. (a) Schematic representation and typical SEM images of guided growth of VLS GaN NWs along the nanofacets on the C-plane sapphire substrates with (1, 2) miscut angle and (3, 4) annealed without miscut angle. (b) Guided growth of VLS Si NW on Si (111) substrate with miscut angle of 10°. (c) Schematic representation of Au droplets

pinned on the hilltop or valley bottom of the 10° miscutted Si (111) surface. (d) Statistical studies on the in-plane growth yield in relationship of miscut angle (4° and 10°) and the disilane pressure. (e) In-plane VLS growth of GaAs NWs on GaAs substrate by Au droplets at 460 °C in a MOCVD system, (f) whose yield is significantly improved up to 100 % by tuning the V/III ratio with NW bottom width higher than 70 nm. (g) Surface structures of epitaxial GaAs grown on GaAs (110) substrates at 450 °C in a MBE system, by varying the V/III ratios and the epilayer growth rates: (1) V/III ratio of 3 and 0.1 μ/h, (2) V/III ratio of 20 and 0.1 μ/h, (3) V/III ratio of 3 and 0.5 μ/h, (4) V/III ratio of 3 and 1 μ/h, (5) V/III ratio of 10 and 1 μ/h.

The guiding effects do not only rely on the nanofaceted substrate, the step edges on the patterned substrates can also lead the guided NW growth, which has been well demonstrated for the in-plane SLS Si NWs.[116, 195-199] Moreover, by *in situ* generating the In NPs on the sidewall of buried ITO nano-pad by a $H_2$ plasma treatment[187], the positioning and the guided growth of Si NWs can be achieved simultaneously, as seen in Figure 13 (a-1), where the growth of In droplet catalyzed in-plane SLS Si NW starts from the buried ITO nano-pad sidewall (see Figure 13 (a-2)) and stops along the $SiO_2$ guided step edge (see Figure 13 (a-3)). Figure 13 (a-4) shows a much longer step-guided Si NW that crosses the entire SEM images (at least longer than 40 μm). The guiding step edge presents two-fold benefits: on the one hand, similar with the pinning effect by the valleys or hilltops on the miscutted substrate (see Figure 12 (c)), the In droplet preferentially wets the step edge due to the S-L system Gibbs free energy minimization (see the illustration in Figure 13 (b-1,2)); on the other hand, the *c*-Si is preferentially precipitated at the step edge, considering the SLS growth mode as a process of heterogeneous nucleation and crystal growth from solution (see the illustration in Figure 13 (b-3)).[143, 158, 200, 201] Thus, the substrate wettability gradient is built up along the guiding step edge, which drives the guided growth of the NW, as illustrated in Figure 13 (b-5). However, the droplet may leave or climb up the guiding step during this step-guided growth, an example of droplet detachment from the guiding step is shown in Figure 13 (b-6). Xu et al. have discussed the step guiding principles by considering the geometries of the SLS system elsewhere.[197] Here, we would emphasize the effect of the *a*-Si:H coating layer, that is, once an unconformal layer is coated, probably due to a lower deposition rate on the step edge in comparison with that on the substrate[202, 203], an asymmetric wetting (see Figure 7 (b)) will occur so that the In droplet will leave the step edge, as illustrated in Figure 13 (b-4).

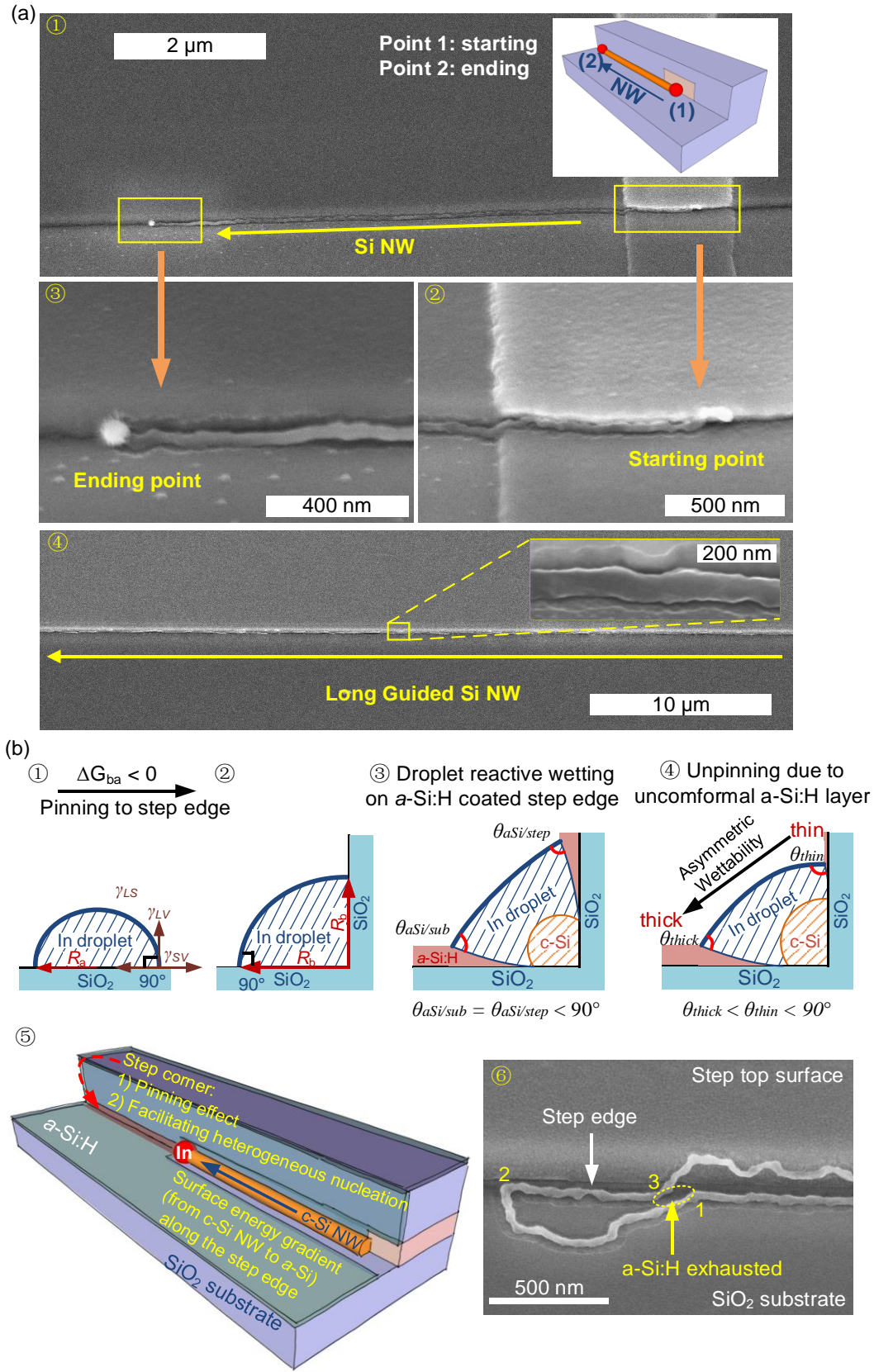

Figure 13. Step-guided growth of in-plane SLS Si NWs on patterned substrates. (a) (1) SEM image of a typical step-guided in-plane SLS Si NWs (2) starting from the buried ITO nanopads and (3) ending

along the step edge. A scheme is illustrated in the inset image. (b) Schematic representation of the evolution of a droplet (1) lying on planar substrate, (2) pinning at the step edge, (3) reactively wetting on the *a*-Si:H coated step edge with *c*-Si precipitation, and (4) the possible unpinning effect due to asymmetric wetting caused by the unconformal *a*-Si:H coating. A step-guided growth mechanism is illustrated in (5), and a failure example of step-guided growth is shown in (6).

## 5. Droplet wetting in the out-of-plane VLS growth process

Figure 10 tells that once the growth of NW pedestal wins the competition against the droplet reactively wetting on the thin film coated substrate surface, the droplet will be lifted up and drive the vertical VLS growth process. However, the droplet wetting behaviors still exist during its out-of-plane transport and profoundly influence the NW morphologies and properties.

### 5.1. Mass loss of droplets due to NW sidewall wetting

The NWs tapering and kinking bring about benefits and troubles depending on their particular applications. Here, we would mainly focus on the NW tapering due to the droplet mass loss during the growth, while the aspect of kinking has been reviewed elsewhere.

The Au surface diffusion on Si NW sidewall are firstly elucidated by the researchers.[204] As shown in Figure 14 (a-1), the Au droplet is completely consumed during the VLS growth of a Si NW at 500 °C, while small Au nano-dots cover the entire sidewall surface in the SEM image. In contrast, when the growth temperature is lowered to 430 °C, the Au can only diffuse in the vicinity of the NW tip, with the droplet partially preserved at the tip (see Figure 14 (a-2)).[205] This can be related to the reactive wetting of Au on Si sidewall surface process (see Figure 3 (a-c)), with the formation of Au nano-dots due to segregation after cooling down. It is expected that lower growth temperature prohibits this reactive wetting behavior. This can be testified by the post growth annealing of Si NWs with Au diffusion. As seen in Figure 14 (b), as long as without the air exposure, the Au can diffuse a long distance from the NW tip to the middle part, no matter annealing in $H_2$ atmosphere (see images 1,2) or in ultra-high vacuum (see images 3, 4). However, once the NWs are exposed in the air, the subsequent annealing in $N_2$ atmosphere can only drive the Au diffusion near the NW top (see images 5, 6).[206]

Apparently, the NW surface oxidation, even in one or two monolayers via a self-passivation manner, insert a barrier between the Au-Si interface and subsequently hinder the Au reactive wetting. This hypothesis is well consistent with the *in situ* TEM observation by Ross el al. As seen in Figure 14 (c), the droplet can preserve its volume when the $O_2$ gas is introduced during the NW growth.[207] In addition, we propose that the oxygen mainly contributes to the NW surface oxidation, as the NW is still the *c*-Si phase rather than $SiO_2$ and the Au screens the oxygen incorporation into the NW from the droplet-NW interface.

The temperature window seems not very wide since in a CVD process a critical thermal energy is mandatory for the decomposition of gaseous precursors and for the overcome of nucleation barrier. Thus, tuning the gas pressure turns to be a more effective means in the vapor deposition process. Indeed, the amount of Si precursors' supplies also affects the Au surface diffusion. The statistical study in Figure 14 (a-3) indicates that the onset of the Au surface diffusion is triggered when lowering the $SiH_4$ partial pressure.[205] A more detailed study provides the dependence of the NW growth rate on the $Si_2H_6$ partial pressure, by varying the $H_2$ partial pressure with the fixed $Si_2H_6$ partial pressure (see Figure 14 (b-8)), or by varying the partial pressure of the $Si_2H_6$ diluted in $H_2$ with fixed total pressures (see Figure 14 (b-9)).[206] Both of the two sets clearly unveil that the low Si NW growth rate with low $Si_2H_6$ partial pressure promotes the Au surface diffusion. Interestingly, when the supply of Si radicals is totally cutted off and then re-provided after a while, the Si nanotrees will grow out from the NW sidewall by the segregated Au nano-dots.[208] This could be regarded as a competition between the Au upward climbing with the downward wetting. As it is difficult to totally cancelled the Au wetting on NW sidewall, elevating the droplet fast tends to be the most practical solution to suppress its surface diffusion to the largest extent.

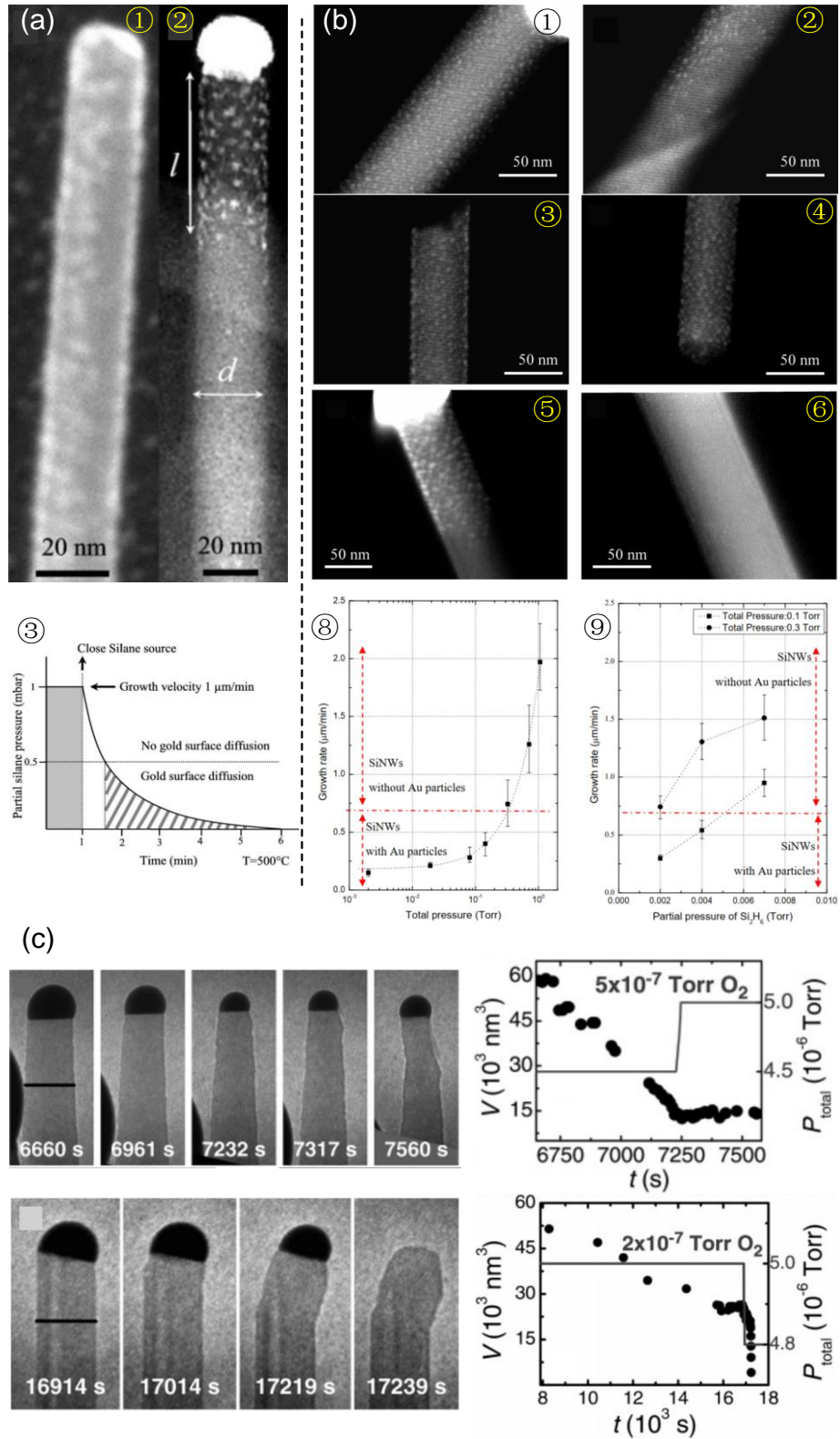

Figure 14. Au droplet surface diffusion on Si NW sidewall. (a) Comparison of Au surface diffusion during NW growth at (1) 500 °C (left) and (2) 430 °C. (3) Statistical analysis of the Au surface diffusion as a function of $SiH_4$ partial pressure in $H_2$. (b) Au droplet surface migration on the top and middle of a Si NW by post-growth annealing in (1, 2) $H_2$ of 0.3 Torr without air exposure, (3, 4) under ultra-high vacuum ($10^{-9}$ Torr) and (5, 6) in $N_2$ atmosphere (1 atm) after air exposure. Statistical studies of the Si NW growth rates as a function (8) of total pressure with fixed $Si_2H_6$ partial pressure of 0.002 Torr by varying the $H_2$ partial pressure, and (9) of the $Si_2H_6$ partial pressure by varying $H_2$ partial pressure for fixed total pressure, showing that low Si supplies favor the Au surface diffusion. (c) *In situ* TEM observation of the evolution of Au droplet volume change by (upper row) introducing and (lower row) switching off $O_2$ gas, showing that Au droplet surface diffusion and the subsequent NW tapering can be hindered by introducing $O_2$ during growth.

The post transition metals such as Sn is an alternative catalyst candidates instead of Au which avoids introducing deep energy impurities that degrade the NW electronic properties.[187, 209-212] Figure 15 (a) displays a systematic study of Sn catalyzed Si NWs growth in a PECVD process. The optimized growth process is towards the high temperature and the diluted $SiH_4$ in $H_2$. This is completely an opposite way against the CVD process for Si NWs growth (see Figure 14).[210] This could be attributed to the difference between the PECVD and CVD processes. Unlike the CVD process where thermal energy contributes to all the steps from the gas decomposition, the radical surface diffusion to the nucleation and the thin film growth, the gas decomposition is activated electrically by the RF power oscillation while the substrate temperature mainly contributes to the rest steps. Most importantly, the Si thin film, very probably in the structure of *a*-Si:H [212], grows faster at lower temperature in higher $SiH_4$ partial pressure atmosphere.[213-220] This helps interpreting the Sn wetting layer during the Si NW as illustrated in Figure 15 (b).[211] Based on the discussion in Section 4.1, Sn will inevitably reactively wet the *a*-Si:H coating layer on the NW sidewall. More importantly, a thicker *a*-Si:H will cause a stronger stretching of the droplet (see Figure 7 (b)), thus the Si NWs grow short and broad at 300 °C in pure $SiH_4$ (see Figure 15 (a)). The droplet sidewall wetting mechanism during a PECVD process is illustrated in Figure 15 (c). We propose that such a sidewall wetting phenomenon can be avoided by oxidizing the NW surface (like that in Figure 14 (c)) or by depositing *μc*-Si:H instead of *a*-Si:H since the interaction between Sn and *μc*-Si:H can be neglected like that of In (see Figure 4 (d)).

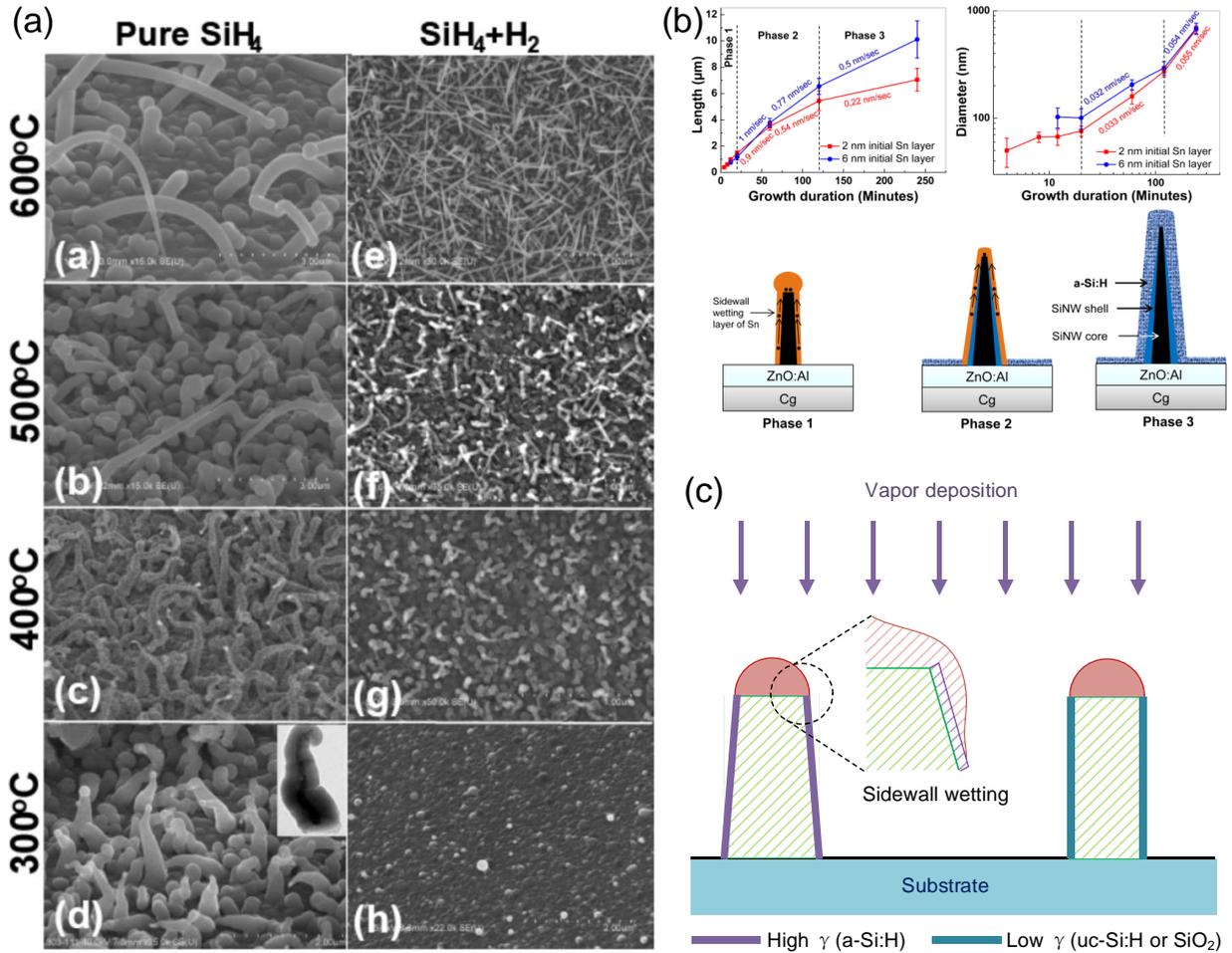

Figure 15. Sn droplet sidewall wetting during VLS Si NW growth in PECVD system. (a) Systematic study of the effects of the silane partial pressure and the growth temperature on the NW morphologies. (b) Statistical studies of (upper left) the NW length as function of growth duration and (upper right) the NW diameter as a function of growth duration, with a scheme of Sn wetting in three phases during NW growth. (c) Schematic representation of a possible Sn sidewall wetting mechanism due to reactive wetting (left), and the possible ways of eliminating the Sn wetting behavior by oxidizing the sidewall surface (correlated with Figure 14 (c)) or by depositing $\mu c$-Si:H instead of $a$-Si:H during VLS growth.

Different with the $a$-Si:H coating layer induced droplet surface wetting, the III-V tapering is mainly governed by the V/III ratios. Fundamentally, the low V partial pressure eliminates the III-V thin film deposition on the NW sidewall, thus the sidewall strain formation can be decayed to the largest extent and the sidewall surface energy is therefore maintained in a low level during NW growth. As a result, the droplet sidewall wetting is suppressed. In particular in a self-catalyzed growth process by the group-III droplet, low V/III ratios may lead a NW broader at the top while smaller at the bottom, since the droplet volume keeps increasing due the excess supplement from the high partial pressure III flux, which subsequently enlarges the droplet-NW interfacial area. Reversely, high V/III deposition will turn the NW sidewall surface with increasing strain energy, which drives the droplet surface diffusion until a complete consumption and leaves the NW a tapering morphology. Figure 16 (a) illustrates such an evolution of droplet volume from low to high V/III ratios during a self-catalyzed growth of GaP NW growth (upper), based on the experimental and theoretical studies on the relationship between NW radius and its corresponding position along the NW.[221] Moreover, Tersoff el al. study the evolution of the droplet shape by varying the III/V ratios, as seen in Figure 16 (b).[222]

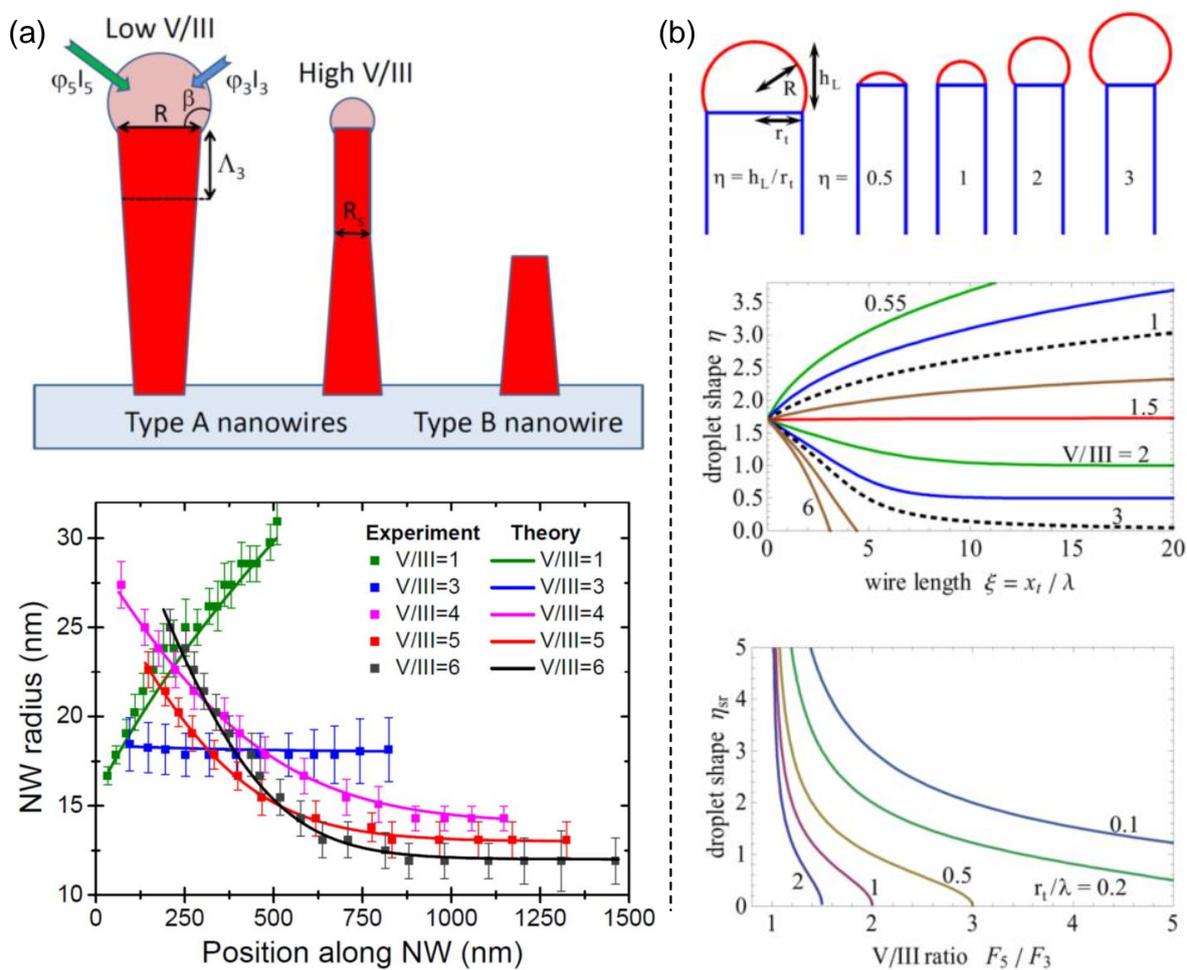

Figure 16. Control of the droplet sidewall wetting by tuning V/III ratio. (a) Schematic representation of droplet volume and the corresponding NW morphologies with low and high V/III ratios (upper) and the statistical analysis of the NW radius as a function of the position along NW, which fits the theoretical calculations (lower). (b) Theoretical studies of the droplet shape evolution during growth for fixed NW diameter (middle) by varying V/III ratios and the steady-state droplet shape versus V/III ratio for fixed NW diameter (lower), with the geometric parameters of the droplet-NW system illustrated in the upper row.

Even though the droplet diffusion induced NW tapering that does not favor their electronic and photonic applications[223], the NW doping profiles can be ameliorated by the wetting layer. For an Au catalyzed Si NW without wetting layer, the heavier doping of phosphorous is observed at the NW bottom and on the NW surface, which gradually attenuates along the NW axis, as seen in the atomic tomographic mapping in Figure 17 (a-1~3).[224] This is because the NW tapering is mainly contributed to the sidewall deposition of Si thin film, where the doping contribution by the vapor-solid (VS) process is higher than the VLS process, as illustrated in Figure 17 (a-4).[225] However, once the droplet sidewall wetting dominates the NW tapering, in particular for those with incorporated catalyst atoms as dopants, the homogeneous doping profiles turn to be possible. Figure 17 (b) shows the STEM-EDX mapping of the Ge (in pink) and Sn (in green) elements in two types of GeSn NWs. For the former one catalyzed by a relatively larger Sn droplet (see image 1-4), disordered Sn clusters are randomly distributed on the tapered NW surface. However, for a needle like GeSn NW with totally consumed Sn catalyst, a homogeneous distribution of Sn is found on the NW surface as well as within the NW. Similarly, the VLS grown Si NW by the InSn alloy catalyst also acquires a very uniform doping concentration, as seen in the atomic tomographic mapping in Figure 17 (c) [212], whose growth experiences the three phases as illustrated in Figure 15 (b).

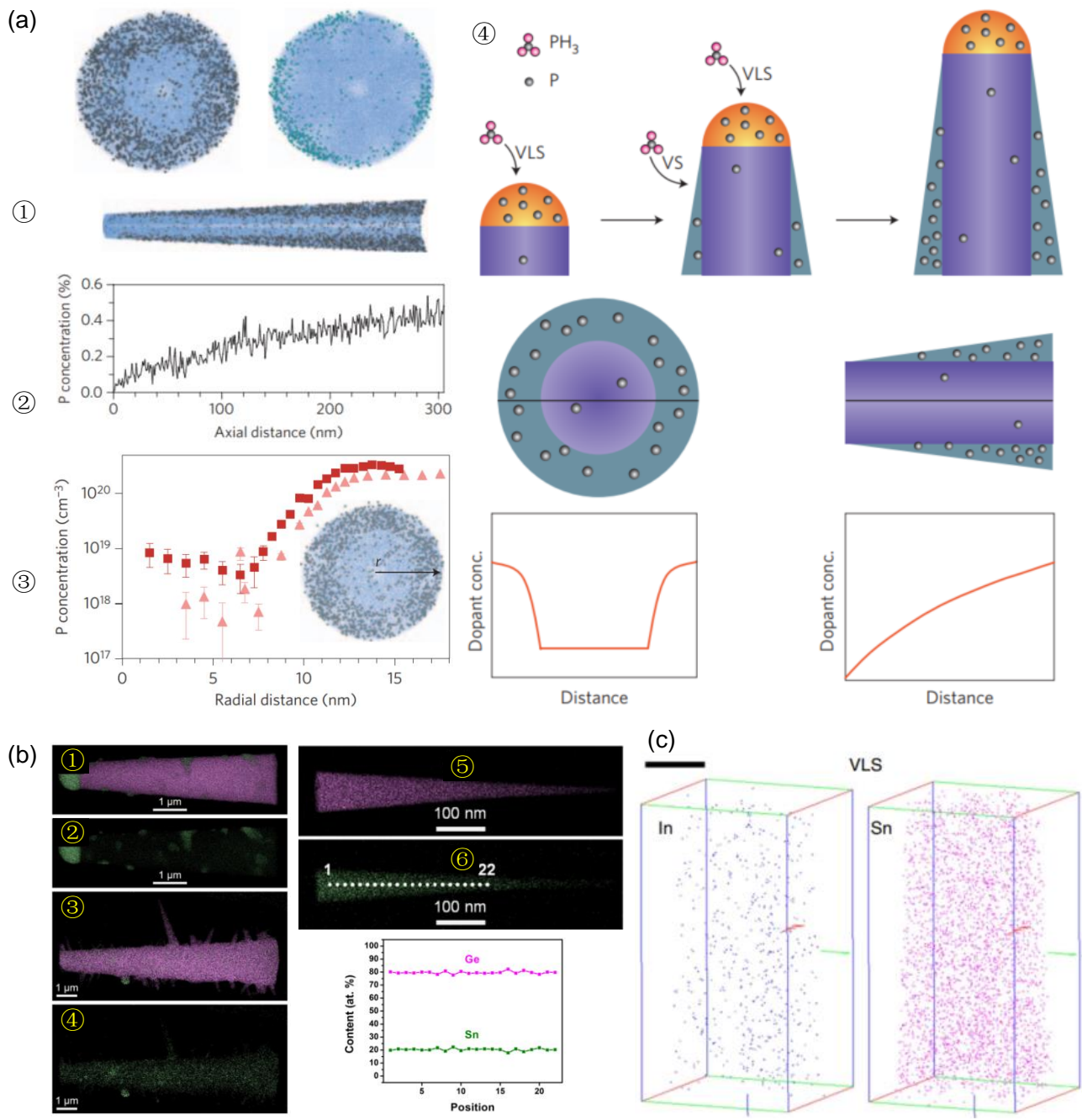

Figure 17. Doping profiles of VLS grown NWs. (a) (1) Distribution of phosphorus (grey dots) and oxygen (blue dots) in a VLS Si NW by atomic tomographic reconstruction technique, with phosphorus distribution along the NW (2) axial distance and (3) radial distance. (4) Schematic representation of the doping routes via VLS process on NW top and via VS process on the NW sidewall. (b) STEM-EDX Ge (in pink) and Sn (in green) mapping of VLS grown GeSn NWs: (1-4) shows the Sn surface diffusion on the sidewall of a gross NW with relatively large Sn droplet on tip, while (5-6) homogeneous Sn incorporation is observed in a small NW with Sn droplet totally consumed. (c)

Homogeneous In and Sn incorporation in VLS grown Si NW catalyzed by InSn alloy droplet.

5.2. Evolution of droplet-NW interface during III-V NWs phase switching

Another important issue of the III-V NWs is the Zinc-Blend-to-Wurtzite (ZB-WZ) phase switching[97], during which the droplet catalyst shows different wetting behaviors. Even though the ZB structure prevails in the VLS-grown III-V NWs, the NW structure switching between WZ and ZB phases are often observed. The TEM image in Figure 18 (a) shows that the GaAs NW grows in ZB phase at the initial stage and rapidly switch to WZ phase. Glas et al. propose that since it takes time for accommodating III/V atoms, the Ga (as well as As) supersaturation in the Au droplet is relatively lower at the initial stage, in comparison with that during the steady state of growth. Thus, it is believed that the WZ growth needs a higher supersaturation, which is favored by a high V/III ratio growth.[226] This interpretation is well demonstrated by the *in situ* TEM observation.[138] Figure 18 (c-1) displays a sequence of TEM images extracted from a dark-field video recording the ZB GaAs layers (bright planes) with AsH$_3$ pressure of $1\times10^{-7}$ Torr and TMGa pressure of $2\times10^{-8}$ Torr. Since the nucleation occurs in the center of the droplet-NW interface, the truncation is formed at the TPL with the droplet dewetting. Subsequently, the growth of the ZB phase proceeds in a periodic truncation filling and jumping fashion. However, the truncation disappears after increasing the AsH$_3$ pressure to $1\times10^{-5}$ Torr, with a steady growth of WZ phase in a step flow manner (see Figure 18 (c-2), the arrows indicate the position of the flowing step). A clearer observation of the step flow growth process is realized by a bird view *in situ* TEM technique, where the nucleation originates from the TPL and the Au-NW contact angle is close to 90 ° (Figure 18 (e)). Moreover, the abruptly ZB-WZ switching can be controllably manipulated by varying the V/III ratio, resulting in the formation of ZB-WZ superlattice, as seen in TEM images in Figure 18 (d) (the growth of the NW starts from the lower image and ends in the upper image, the bright planes indicates the ZB phase and the dark segments indicates the WZ phase). Importantly, the geometric measurement of the droplet aspect ratios during ZB-WZ switching implies that droplet wets the WZ surface while dewets the ZB surface, as illustrated in Figure 18 (b).

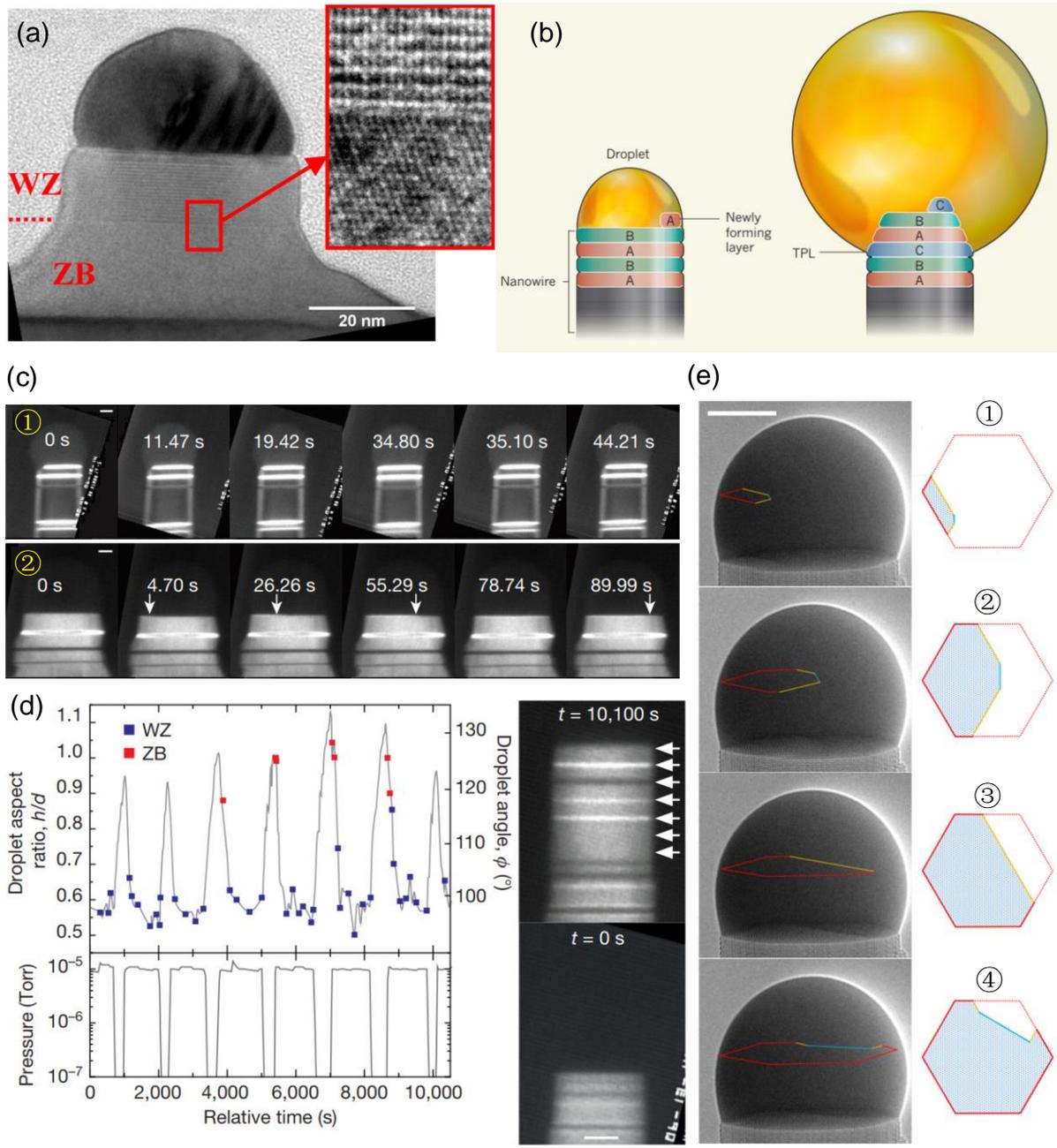

Figure 18. WZ-ZB phase transition in III-V NWs. (a) A TEM image of ZB-WZ stacking in Au catalyzed VLS grown GaAs NW. (b) Illustration of different droplet-NW interfaces of WZ (left) and ZB (right) III-V NWs, where an ABAB stacking growth process of a WZ NW is acquired when the droplet contact angle is close to 90°, whereas a ZB NW will grow with a larger or smaller droplet contact angle. (c) *In situ* TEM observation of ZB (upper) and WZ (lower) GaAs by Au droplet. (d) The WZ-to-ZB phase transition can be achieved by decreasing the As partial pressure. (e) In situ TEM observation of the atomic step flow initiating at the TPL during a WZ GaAs growth process, with the Au droplet contact angle close to 90°.

# 6. Droplet surface migration for producing 2D nanoribbons

In the last section, we would discuss the formation of 2D NRs by droplet surface migration. The mechanism, no matter either in an etching manner or in a growth manner, meets well with the theory of reactive wetting induced droplet spontaneous motion by de Gennes.[44]

6.1. Tailoring graphene nanoribbons by crawling droplets

The CVD process prevails the large-scale production of graphene thin films, which relies on the carbon deposition on catalytic metal foils. The growth on the Cu foils is governed by a surface reaction mechanism due to the super low solubility of carbon in Cu. In contrast, a much larger amount of carbon can be dissolved in to the Ni foils, so the graphene thin films are grown in a carbon segregation and precipitation process. Interestingly, the latter growth mode can proceed reversely, that is, a Ni droplet is able to move on a graphene thin film and dissolve the graphene in its path, as illustrated in Figure 19 (a).[227, 228] Thus, the graphene NRs will be formed once two Ni droplets move in parallel, as seen Figure 19 (b, c). It is expected that the width of NRs will be precisely designed once the spacing of Ni NPs are controllable defined.

6.2. $MoS_2$ nanoribbons growth via vapor-solid-liquid process

Recently, a direct growth of $MoS_2$ NRs has been realized via a VLS process.[229] Figure 19 (d-1,2) illustrate a large view of the growth process. The growth process is a simple epitaxial growth $MoS_2$ on NaCl substrate in a MOCVD system. However, besides the growth of the conventional triangular sheets, the straight or kinked NRs are also produced, which are led by the Na-Cl-O droplets formed by the reaction between the sublimated Na and Cl atoms from the substrates and O atoms from $MoO_3$ precursors (see the optical image in Figure 19 (e-1)). A focus on a single NR under SEM unveils the boundary of a droplet moving path, with NRs in the center (see Figure 19 (e-2)). This is very similar with the trench of the in-plane SLS process (see Figure 4 (a)). Moreover, the moving droplet leaves a fish-bone like substrate surface along its path, as seen in Figure 19 (e-3). Furthermore, the mass loss during the droplet motion (see the tiny dots along the edges of the NRs in Figure 19 (e-4,5)) and the self-avoiding growth (see the zigzag NR between two parallel NRs in Figure 19 (e-6)) provides more similarities with the in-plane SLS growth process. Last, the growth mechanism as illustrated in Figure 19 (f) meets well with the interpretation of in-plane SLS growth (see Figure 7 (g)), that is, the droplet dewets the substrate once $MoS_2$ is precipitated underneath it, thereby its surface migration is activated along the wettability gradient.

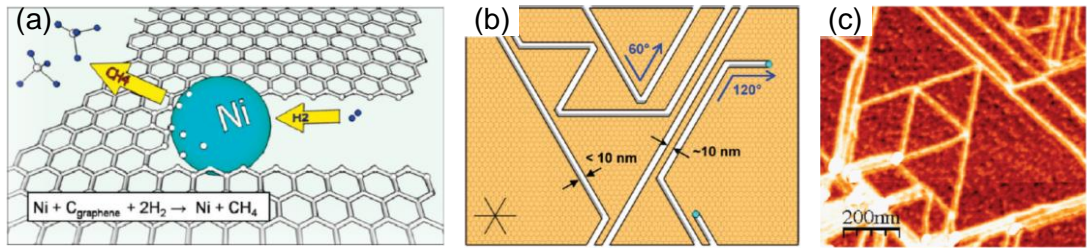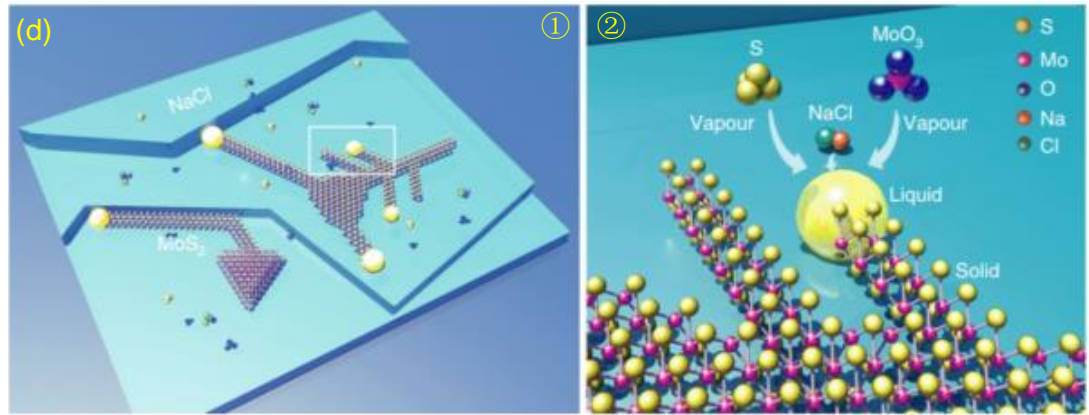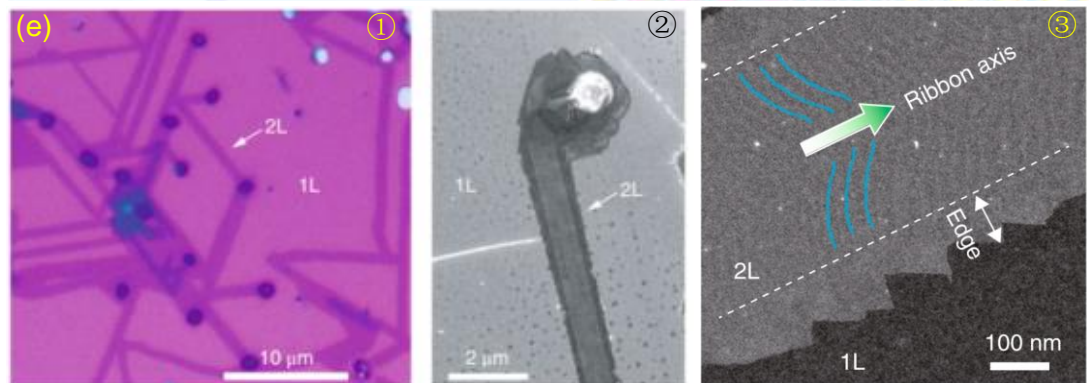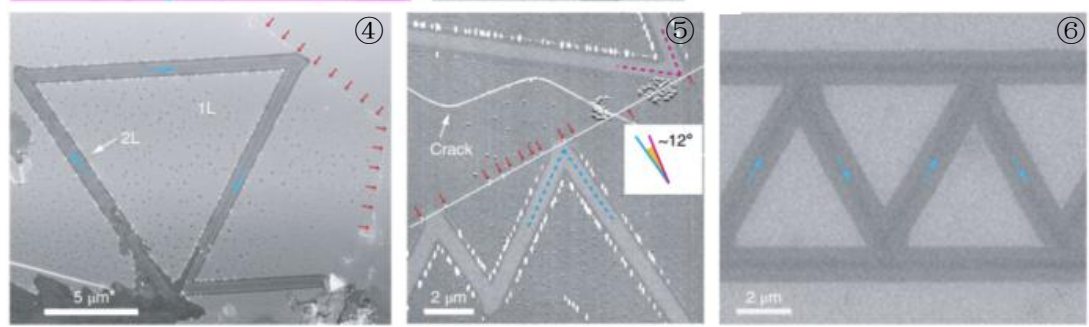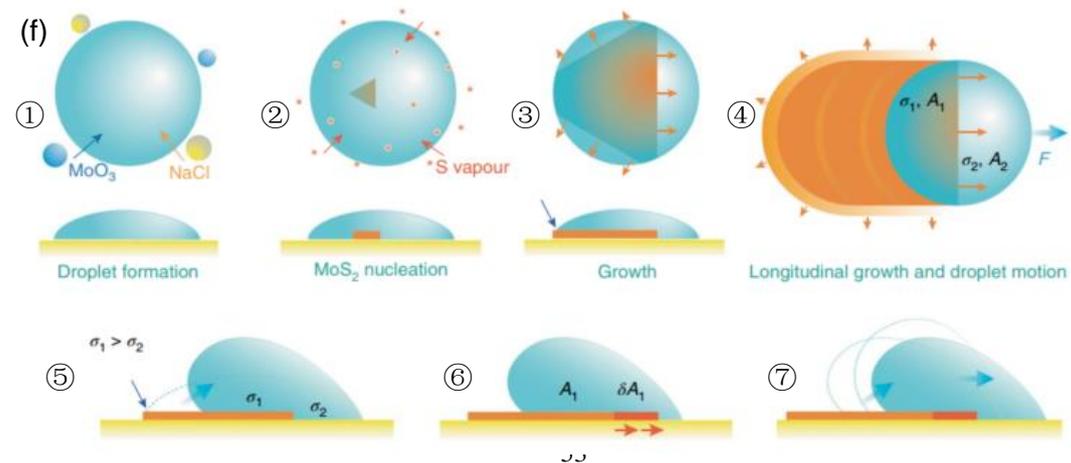

Figure 19. Metal droplet surface migration for producing 2D nanoribbons. Graphene NRs by mobile droplet etching: (a) Mechanism of the graphene etching by Ni droplet. (b) Schematic representation of migrating Ni droplet for tailoring the graphene sheet into various shapes involving NRs, with an AFM image in (c). $MoS_2$ NRs growth by spontaneous droplet migration: (d) schematic representation of the VLS growth of $MoS_2$ NRs catalyzed by Na-Cl-O droplet migration on NaCl substrate. (e) Microscopic observation of $MoS_2$ NRs: (1) optical images of NRs on a monolayer $MoS_2$ film; (2) SEM image of a NR located in the trench (see the NR/trench contrast) with a Na-Cl-O NP at its end; (3) STEM image shows the zigzag edges and fishbone-lie contrast at the center of the NR; (4, 5) SEM and AFM images of self-turning NRs, with tiny NPs along their paths; (6) a zigzag shaped NR whose growth path is confined within two parallel NRs. (f) Schematic representation of the mechanism of in-plane VLS growth of $MoS_2$ NRs: the droplet dewets the substrate once $MoS_2$ is precipitated underneath it, thereby its surface migration is activated along the wettability gradient, in a very similar manner as illustrated in Figure 7.

## 7. Conclusion

To summary, we review the low dimensional nanomaterials growth mediated by catalytic metal droplets, in a viewpoint of dynamic droplet evolution under vapor phase deposition. We summarize several key factors that affect the droplet spreading behaviors and their consequent nanofluidic transport, which involves deposition parameters, solid-liquid interfaces, crystal phases, substrate nanofacets and so on, which deterministically results in various morphologies and growth directions of the nanomaterials.

## Reference

[1] A.P. Alivisatos, Semiconductor Clusters, Nanocrystals, and Quantum Dots, Science, 271 (1996) 933-937.

[2] A.P. Alivisatos, Perspectives on the Physical Chemistry of Semiconductor Nanocrystals, The Journal of Physical Chemistry, 100 (1996) 13226-13239.

[3] J. Hu, T.W. Odom, C.M. Lieber, Chemistry and Physics in One Dimension: Synthesis and Properties of Nanowires and Nanotubes, Accounts of Chemical Research, 32 (1999) 435-445.

[4] M. Law, J. Goldberger, P. Yang, SEMICONDUCTOR NANOWIRES AND NANOTUBES, Annual Review of Materials Research, 34 (2004) 83-122.

[5] Y. Li, F. Qian, J. Xiang, C.M. Lieber, Nanowire electronic and optoelectronic devices, Materials Today, 9 (2006) 18-27.

[6] O. Hayden, R. Agarwal, W. Lu, Semiconductor nanowire devices, Nano Today, 3 (2008) 12-22.

[7] P. Yang, R. Yan, M. Fardy, Semiconductor Nanowire: What's Next?, Nano Letters, 10 (2010) 1529-1536.

[8] V. Schmidt, J.V. Wittemann, S. Senz, U. Gösele, Silicon Nanowires: A Review on Aspects of their Growth and their Electrical Properties, Advanced Materials, 21 (2009) 2681-2702.

[9] N.P. Dasgupta, J. Sun, C. Liu, S. Brittman, S.C. Andrews, J. Lim, H. Gao, R. Yan, P. Yang, 25th Anniversary Article: Semiconductor Nanowires – Synthesis, Characterization, and Applications, Advanced Materials, 26 (2014) 2137-2184.

[10] Y. Wang, T. Wang, P. Da, M. Xu, H. Wu, G. Zheng, Silicon Nanowires for Biosensing, Energy Storage, and Conversion, Advanced Materials, (2013) 5177–5195.

[11] L. Guniat, P. Caroff, I.M.A. Fontcuberta, Vapor Phase Growth of Semiconductor Nanowires: Key Developments and Open Questions, Chem Rev, 119 (2019) 8958-8971.

[12] X. Wang, J. Song, F. Zhang, C. He, Z. Hu, Z. Wang, Electricity Generation based on One-Dimensional Group-III Nitride Nanomaterials, Advanced Materials, 22 (2010) 2155-2158.

[13] H. Yan, H.S. Choe, S. Nam, Y. Hu, S. Das, J.F. Klemic, J.C. Ellenbogen, C.M. Lieber, Programmable nanowire circuits for nanoprocessors, Nature, 470 (2011) 240-244.

[14] B. Tian, X. Zheng, T.J. Kempa, Y. Fang, N. Yu, G. Yu, J. Huang, C.M. Lieber, Coaxial silicon nanowires as solar cells and nanoelectronic power sources, Nature, 449 (2007) 885-889.

[15] A.I. Boukai, Y. Bunimovich, J. Tahir-Kheli, J.-K. Yu, W.A. Goddard III, J.R. Heath, Silicon nanowires as efficient thermoelectric materials, Nature, 451 (2008).

[16] M.S. Gudiksen, L.J. Lauhon, J. Wang, D.C. Smith, C.M. Lieber, Growth of nanowire superlattice structures for nanoscale photonics and electronics, Nature, 415 (2002).

[17] R. Yan, D. Gargas, P. Yang, Nanowire photonics, Nat Photon, 3 (2009) 569-576.

[18] R. He, P. Yang, Giant piezoresistance effect in silicon nanowires, Nat Nano, 1 (2006).

[19] Y. Cui, C.M. Lieber, Functional Nanoscale Electronic Devices Assembled Using Silicon Nanowire Building Blocks, Science, 291 (2001) 851-853.

[20] Y. Cui, Q. Wei, H. Park, C.M. Lieber, Nanowire Nanosensors for Highly Sensitive and Selective Detection of Biological and Chemical Species, Science, 293 (2001) 1289-1292.

[21] K.-I. Chen, B.-R. Li, Y.-T. Chen, Silicon nanowire field-effect transistor-based biosensors for biomedical diagnosis and cellular recording investigation, Nano Today, 6 (2011) 131-154.

[22] R.J. Warburton, C. Schäflein, D. Haft, F. Bickel, A. Lorke, K. Karrai, J.M. Garcia, W. Schoenfeld, P.M. Petroff, Optical emission from a charge-tunable quantum ring, Nature, 405 (2000) 926-929.

[23] K. Karrai, R.J. Warburton, C. Schulhauser, A. Högele, B. Urbaszek, E.J. McGhee, A.O. Govorov, J.M. Garcia, B.D. Gerardot, P.M. Petroff, Hybridization of electronic states in quantum dots through photon emission, Nature, 427 (2004) 135-138.

[24] B.D. Gerardot, D. Brunner, P.A. Dalgarno, P. Öhberg, S. Seidl, M. Kroner, K. Karrai, N.G. Stoltz, P.M. Petroff, R.J. Warburton, Optical pumping of a single hole spin in a quantum dot, Nature, 451 (2008) 441-444.

[25] K.H. Schmidt, G. Medeiros-Ribeiro, J. Garcia, P.M. Petroff, Size quantization effects in InAs self-assembled quantum dots, Applied Physics Letters, 70 (1997) 1727-1729.

[26] H. Zhang, D.E. Liu, M. Wimmer, L.P. Kouwenhoven, Next steps of quantum transport in Majorana nanowire devices, Nat Commun, 10 (2019) 5128.

[27] S. Yip, L. Shen, J.C. Ho, Recent advances in III-Sb nanowires: from synthesis to applications, Nanotechnology, 30 (2019) 202003.

[28] J. Wong-Leung, I. Yang, Z. Li, S.K. Karuturi, L. Fu, H.H. Tan, C. Jagadish, Engineering III-V Semiconductor Nanowires for Device Applications, Adv Mater, (2019) e1904359.

[29] C. Jia, Z. Lin, Y. Huang, X. Duan, Nanowire Electronics: From Nanoscale to Macroscale, Chem Rev, 119 (2019) 9074-9135.

[30] P.G. de Gennes, Wetting: statics and dynamics, Reviews of Modern Physics, 57 (1985) 827-863.

[31] J.M. Howe, Bonding, structure, and properties of metal/ceramic interfaces: Part 1 Chemical bonding, chemical reaction, and interfacial structure, International Materials Reviews, 38 (1993) 233-256.

[32] J.M. Howe, Bonding, structure, and properties of metal/ceramic interfaces: Part 2 Interface fracture behaviour and property measurement, International Materials Reviews, 38 (1993) 257-271.

[33] J. De Coninck, T.D. Blake, Wetting and Molecular Dynamics Simulations of Simple Liquids, Annual Review of Materials Research, 38 (2008) 1-22.

[34] D.R. Heine, G.S. Grest, E.B. Webb, Surface Wetting of Liquid Nanodroplets:

Droplet-Size Effects, Physical Review Letters, 95 (2005) 107801.

[35] W. Mönch, Metal-semiconductor contacts: electronic properties, Surface Science, 299–300 (1994) 928-944.

[36] A.S. Skapski, The Surface Tension of Liquid Metals, The Journal of Chemical Physics, 16 (1948) 389-393.

[37] N.K. Adam, H.K. Livingston, Contact Angles and Work of Adhesion, Nature, 182 (1958) 128-128.

[38] D. Seveno, T.D. Blake, J.l. De Coninck, Young's Equation at the Nanoscale, Physical Review Letters, 111 (2013) 096101.

[39] N. Sobczak, M. Singh, R. Asthana, High-temperature wettability measurements in metal/ceramic systems: Some methodological issues, Current Opinion in Solid State and Materials Science, 9 (2005) 241-253.

[40] D. Chatain, Anisotropy of Wetting, Annual Review of Materials Research, 38 (2008) 45-70.

[41] G. Kumar, K.N. Prabhu, Review of non-reactive and reactive wetting of liquids on surfaces, Advances in Colloid and Interface Science, 133 (2007) 61-89.

[42] H.P. Greenspan, On the motion of a small viscous droplet that wets a surface, Journal of Fluid Mechanics, 84 (1978) 125-143.

[43] F. Brochard, Motions of droplets on solid surfaces induced by chemical or thermal gradients, Langmuir, 5 (1989) 432-438.

[44] P.G. de Gennes, The dynamics of reactive wetting on solid surfaces, Physica A: Statistical Mechanics and its Applications, 249 (1998) 196-205.

[45] K. Ichimura, S.-K. Oh, M. Nakagawa, Light-Driven Motion of Liquids on a Photoresponsive Surface, Science, 288 (2000) 1624-1626.

[46] F.D. Dos Santos, T. Ondarçuhu, Free-Running Droplets, Physical Review Letters, 75 (1995) 2972-2975.

[47] M.K. Chaudhury, G.M. Whitesides, How to Make Water Run Uphill, Science, 256 (1992) 1539-1541.

[48] B.S. Gallardo, V.K. Gupta, F.D. Eagerton, L.I. Jong, V.S. Craig, R.R. Shah, N.L. Abbott, Electrochemical Principles for Active Control of Liquids on Submillimeter Scales, Science, 283 (1999) 57-60.

[49] Y. Sumino, N. Magome, T. Hamada, K. Yoshikawa, Self-Running Droplet: Emergence of Regular Motion from Nonequilibrium Noise, Physical Review Letters, 94 (2005) 068301.

[50] A.K. Schmid, N.C. Bartelt, R.Q. Hwang, Alloying at Surfaces by the Migration of Reactive Two-Dimensional Islands, Science, 290 (2000) 1561-1564.

[51] J. Tersoff, D.E. Jesson, W.X. Tang, Running Droplets of Gallium from Evaporation of Gallium Arsenide, Science, 324 (2009) 236-238.

[52] T.R. Anthony, H.E. Cline, Thermomigration of Gold-Rich Droplets in Silicon, Journal of Applied Physics, 43 (1972) 2473-2476.

[53] A. Moosavi, M. Rauscher, S. Dietrich, Motion of Nanodroplets near Edges and Wedges, Physical Review Letters, 97 (2006) 236101.

[54] H.A. Stone, A.D. Stroock, A. Ajdari, Engineering Flows in Small Devices, Annual Review of Fluid Mechanics, 36 (2004) 381-411.

[55] L. Feng, S. Li, Y. Li, H. Li, L. Zhang, J. Zhai, Y. Song, B. Liu, L. Jiang, D. Zhu, Super-Hydrophobic Surfaces: From Natural to Artificial, Advanced Materials, 14 (2002) 1857-1860.

[56] C. Pei-Yu, C. Zehao, M.C. Wu, Droplet Manipulation With Light on Optoelectrowetting Device, Journal of Microelectromechanical Systems, 17 (2008) 133-138.

[57] J. Hong, J.B. Edel, A.J. deMello, Micro- and nanofluidic systems for high-throughput biological screening, Drug Discov Today, 14 (2009) 134-146.

[58] Y. Zheng, H. Bai, Z. Huang, X. Tian, F.-Q. Nie, Y. Zhao, J. Zhai, L. Jiang, Directional water collection on wetted spider silk, Nature, 463 (2010) 640-643.

[59] C. Liu, J. Ju, J. Ma, Y. Zheng, L. Jiang, Directional drop transport achieved on high-temperature anisotropic wetting surfaces, Adv Mater, 26 (2014) 6086-6091.

[60] R. Seemann, M. Brinkmann, T. Pfohl, S. Herminghaus, Droplet based microfluidics, Rep Prog Phys, 75 (2012) 016601.

[61] Q. Zhao, H. Cui, Y. Wang, X. Du, Microfluidic Platforms toward Rational Material Fabrication for Biomedical Applications, Small, (2019) e1903798.

[62] Q. Sun, D. Wang, Y. Li, J. Zhang, S. Ye, J. Cui, L. Chen, Z. Wang, H.J. Butt, D. Vollmer, X. Deng, Surface charge printing for programmed droplet transport, Nat Mater, 18 (2019) 936-941.

[63] C. Liu, J. Sun, Y. Zhuang, J. Wei, J. Li, L. Dong, D. Yan, A. Hu, X. Zhou, Z. Wang, Self-propelled droplet-based electricity generation, Nanoscale, 10 (2018) 23164-23169.

[64] I.A. Aksay, C.E. Hoge, J.A. Pask, Wetting under chemical equilibrium and nonequilibrium conditions, The Journal of Physical Chemistry, 78 (1974) 1178-1183.

[65] N. Eustathopoulos, B. Drevet, Interfacial bonding, wettability and reactivity in metal/oxide systems, J. Phys. III France, 4 (1994) 1865-1881.

[66] D. Chatain, L. Coudurier, N. Eustathopoulos, Wetting and interfacial bonding in ionocovalent oxide-liquid metal systems, Rev. Phys. Appl. (Paris), 23 (1988) 1055-1064.

[67] N. Eustathopoulos, D. Chatain, L. Coudurier, Wetting and interfacial chemistry in liquid metal-ceramic systems, Materials Science and Engineering: A, 135 (1991) 83-88.

[68] K. Landry, C. Rado, N. Eustathopoulos, Influence of interfacial reaction

rates on the wetting driving force in metal/ceramic systems, Metallurgical and Materials Transactions A, 27 (1996) 3181-3186.

[69] N. Eustathopoulos, Dynamics of wetting in reactive metal/ceramic systems, Acta Materialia, 46 (1998) 2319-2327.

[70] C. Rado, S. Kalogeropoulou, N. Eustathopoulos, Bonding and wetting in non-reactive metal/SiC systems: weak or strong interfaces?, Materials Science and Engineering: A, 276 (2000) 195-202.

[71] N. Eustathopoulos, Progress in understanding and modeling reactive wetting of metals on ceramics, Current Opinion in Solid State and Materials Science, 9 (2005) 152-160.

[72] O. Dezellus, N. Eustathopoulos, Fundamental issues of reactive wetting by liquid metals, Journal of materials Science, 45 (2010) 4256-4264.

[73] P. Protsenko, J.P. Garandet, R. Voytovych, N. Eustathopoulos, Thermodynamics and kinetics of dissolutive wetting of Si by liquid Cu, Acta Materialia, 58 (2010) 6565-6574.

[74] L. Yin, A. Chauhan, T.J. Singler, Reactive wetting in metal/metal systems: Dissolutive versus compound-forming systems, Materials Science and Engineering: A, 495 (2008) 80-89.

[75] L. Yin, B.T. Murray, S. Su, Y. Sun, Y. Efraim, H. Taitelbaum, T.J. Singler, Reactive wetting in metal-metal systems, Journal of Physics: Condensed Matter, 21 (2009) 464130.

[76] U. Denker, A. Rastelli, M. Stoffel, J. Tersoff, G. Katsaros, G. Costantini, K. Kern, N.Y. Jin-Phillipp, D.E. Jesson, O.G. Schmidt, Lateral Motion of SiGe Islands Driven by Surface-Mediated Alloying, Physical Review Letters, 94 (2005) 216103.

[77] Y.W. Mo, D.E. Savage, B.S. Swartzentruber, M.G. Lagally, Kinetic pathway in Stranski-Krastanov growth of Ge on Si(001), Physical Review Letters, 65 (1990) 1020-1023.

[78] G.S. Solomon, J.A. Trezza, A.F. Marshall, J.J.S. Harris, Vertically Aligned and Electronically Coupled Growth Induced InAs Islands in GaAs, Physical Review Letters, 76 (1996) 952-955.

[79] B. Daudin, F. Widmann, G. Feuillet, Y. Samson, M. Arlery, J.L. Rouvière, Stranski-Krastanov growth mode during the molecular beam epitaxy of highly strained GaN, Physical Review B, 56 (1997) R7069-R7072.

[80] A. Baskaran, P. Smereka, Mechanisms of Stranski-Krastanov growth, Journal of Applied Physics, 111 (2012) -.

[81] R.B. Lewis, P. Corfdir, H. Li, J. Herranz, C. Pfüller, O. Brandt, L. Geelhaar, Quantum Dot Self-Assembly Driven by a Surfactant-Induced Morphological Instability, Physical Review Letters, 119 (2017) 086101.

[82] K. Jacobi, Atomic structure of InAs quantum dots on GaAs, Progress in

Surface Science, 71 (2003) 185-215.

[83] J.G. Keizer, J. Bocquel, P.M. Koenraad, T. Mano, T. Noda, K. Sakoda, Atomic scale analysis of self assembled GaAs/AlGaAs quantum dots grown by droplet epitaxy, Applied Physics Letters, 96 (2010) 062101.

[84] T. Mano, M. Abbarchi, T. Kuroda, C.A. Mastrandrea, A. Vinattieri, S. Sanguinetti, K. Sakoda, M. Gurioli, Ultra-narrow emission from single GaAs self-assembled quantum dots grown by droplet epitaxy, Nanotechnology, 20 (2009) 395601.

[85] K. Kawasaki, D. Yamazaki, A. Kinoshita, H. Hirayama, K. Tsutsui, Y. Aoyagi, GaN quantum-dot formation by self-assembling droplet epitaxy and application to single-electron transistors, Applied Physics Letters, 79 (2001) 2243-2245.

[86] J.-M. Lee, D. Hyun Kim, H. Hong, J.-C. Woo, S.-J. Park, Growth of InAs nanocrystals on GaAs(100) by droplet epitaxy, Journal of Crystal Growth, 212 (2000) 67-73.

[87] J. Wu, Y. Hirono, X. Li, Z.M. Wang, J. Lee, M. Benamara, S. Luo, Y.I. Mazur, E.S. Kim, G.J. Salamo, Self-Assembly of Multiple Stacked Nanorings by Vertically Correlated Droplet Epitaxy, Advanced Functional Materials, 24 (2014) 530-535.

[88] C. Somaschini, S. Bietti, S. Sanguinetti, N. Koguchi, A. Fedorov, Self-assembled GaAs/AlGaAs coupled quantum ring-disk structures by droplet epitaxy, Nanotechnology, 21 (2010) 125601.

[89] C. Somaschini, S. Bietti, N. Koguchi, S. Sanguinetti, Shape control via surface reconstruction kinetics of droplet epitaxy nanostructures, Applied Physics Letters, 97 (2010) 203109.

[90] X.L. Li, G.W. Yang, Growth Mechanisms of Quantum Ring Self-Assembly upon Droplet Epitaxy, The Journal of Physical Chemistry C, 112 (2008) 7693-7697.

[91] K.A. Sablon, J.H. Lee, Z.M. Wang, J.H. Shultz, G.J. Salamo, Configuration control of quantum dot molecules by droplet epitaxy, Applied Physics Letters, 92 (2008) 203106.

[92] X.L. Li, G.W. Yang, On the physical understanding of quantum rings self-assembly upon droplet epitaxy, Journal of Applied Physics, 105 (2009) 103507.

[93] J.H. Lee, Z.M. Wang, Z.Y. Abuwaar, N.W. Strom, G.J. Salamo, Evolution between self-assembled single and double ring-like nanostructures, Nanotechnology, 17 (2006) 3973-3976.

[94] S. Huang, Z. Niu, Z. Fang, H. Ni, Z. Gong, J. Xia, Complex quantum ring structures formed by droplet epitaxy, Applied Physics Letters, 89 (2006) 031921.

[95] T. Mano, N. Koguchi, Nanometer-scale GaAs ring structure grown by droplet epitaxy, Journal of Crystal Growth, 278 (2005) 108-112.

[96] Z. Gong, Z.C. Niu, S.S. Huang, Z.D. Fang, B.Q. Sun, J.B. Xia, Formation of GaAs∕AlGaAs and InGaAs∕GaAs nanorings by droplet molecular-beam epitaxy, Applied Physics Letters, 87 (2005) 093116.

[97] F. Glas, J.-C. Harmand, G. Patriarche, Why Does Wurtzite Form in Nanowires of III-V Zinc Blende Semiconductors?, Physical Review Letters, 99 (2007) 146101.

[98] V.G. Dubrovskii, N.V. Sibirev, J.C. Harmand, F. Glas, Growth kinetics and crystal structure of semiconductor nanowires, Physical Review B, 78 (2008) 235301.

[99] R.E. Algra, M.A. Verheijen, M.T. Borgström, L.-F. Feiner, G. Immink, W.J.P. van Enckevort, E. Vlieg, E.P.A.M. Bakkers, Twinning superlattices in indium phosphide nanowires, Nature, 456 (2008) 369-372.

[100] P. Spencer, C. C., M. W., B. H., R. D., Triggering InAs/GaAs Quantum Dot nucleation and growth rate determination by in-situ modulation of surface energy, arXiv:1906.05842v1 (2019).

[101] R.P. Mirin, A. Roshko, M.v.d. Puijl, A.G. Norman, Formation of InAs/GaAs quantum dots by dewetting during cooling, Journal of Vacuum Science & Technology B: Microelectronics and Nanometer Structures Processing, Measurement, and Phenomena, 20 (2002) 1489-1492.

[102] T. Mano, T. Kuroda, K. Kuroda, K. Sakoda, Self-assembly of quantum dots and rings by droplet epitaxy and their optical properties, Journal of Nanophotonics, 3 (2009) 031605.

[103] A. Hiraki, A Model on the Mechanism of Room Temperature Interfacial Intermixing Reaction in Various Metal-Semiconductor Couples: What Triggers the Reaction?, Journal of The Electrochemical Society, 127 (1980) 2662-2665.

[104] A. Hiraki, Low Temperature Reactions at Si-Metal Contacts:From SiO2 Growth due to Si-Au Reaction to the Mechanism of Silicide Formation, Jpn. J. Appl. Phys., 22 (1983) 13.

[105] A. Hiraki, Recent developments on metal-silicon interfaces, Applied Surface Science, 56-58, Part 1 (1992) 370-381.

[106] T.J. Konno, R. Sinclair, Metal-contact-induced crystallization of semiconductors, Materials Science and Engineering: A, 179–180, Part 1 (1994) 426-432.

[107] G. Ottaviani, D. Sigurd, V. Marrello, J.O. McCaldin, J.W. Mayer, Crystal Growth of Silicon and Germanium in Metal Films, Science, 180 (1973) 948-949.

[108] A. Hiraki, M. Nicolet, J.W. Mayer, LOW-TEMPERATURE MIGRATION OF SILICON IN THIN LAYERS OF GOLD AND PLATINUM, Applied Physics Letters, 18 (1971) 178-181.

[109] A. Hiraki, E. Lugujjo, Low-Temperature Migration of Silicon in Metal

Films on Silicon Substrates Studied by Backscattering Techniques, Journal of Vacuum Science & Technology, 9 (1972) 155-158.

[110] N. Ferralis, F.E. Gabaly, A.K. Schmid, R. Maboudian, C. Carraro, Real-Time Observation of Reactive Spreading of Gold on Silicon, Physical Review Letters, 103 (2009) 256102.

[111] Y. Wakayama, S.-i. Tanaka, Kinetics of surface droplet epitaxy and its application to fabrication of mushroom-shaped metal/Si heterostructure on nanometer scale, Surface Science, 420 (1999) 190-199.

[112] H. Miura, M. Kamiko, K. Kyuno, Novel Crystallization Process for Germanium Thin Films: Surfactant-Crystallization Method, Japanese Journal of Applied Physics, 52 (2013) 010204.

[113] V. Neimash, V. Poroshin, P. Shepeliavyi, V. Yukhymchuk, V. Melnyk, A. Kuzmich, V. Makara, A.O. Goushcha, Tin induced a-Si crystallization in thin films of Si-Sn alloys, Journal of Applied Physics, 114 (2013) -.

[114] R.S. Wagner, W.C. Ellis, VAPOR-LIQUID-SOLID MECHANISM OF SINGLE CRYSTAL GROWTH, Applied Physics Letters, 4 (1964) 89-90.

[115] R.S. Wagner, W.C. Ellis, The vapor-Liquid-Solid Mechanism of Crystal Growth and Its Application to Si, Transcations of The Metallurgical Society of AIME, 233 (1965) 12.

[116] Y. Sun, T. Dong, L. Yu, J. Xu, K. Chen, Planar Growth, Integration, and Applications of Semiconducting Nanowires, Advanced Materials, n/a (2019) 1903945.

[117] D. Jariwala, T.J. Marks, M.C. Hersam, Mixed-dimensional van der Waals heterostructures, Nature Materials, 16 (2017) 170-181.

[118] E. Mataev, S.K. Rastogi, A. Madhusudan, J. Bone, N. Lamprinakos, Y. Picard, T. Cohen-Karni, Synthesis of Group IV Nanowires on Graphene: The Case of Ge Nanocrawlers, Nano Letters, 16 (2016) 5267-5272.

[119] F.M. Ross, In Situ Transmission Electron Microscopy, in: P.W. Hawkes, J.C.H. Spence (Eds.) Science of Microscopy, Springer New York, New York, NY, 2007, pp. 445-534.

[120] F.M. Ross, Growth processes and phase transformations studied in situ transmission electron microscopy, IBM Journal of Research and Development, 44 (2000) 489-501.

[121] Y. Wu, P. Yang, Direct Observation of Vapor−Liquid−Solid Nanowire Growth, Journal of the American Chemical Society, 123 (2001) 3165-3166.

[122] A.I. Persson, M.W. Larsson, S. Stenstrom, B.J. Ohlsson, L. Samuelson, L.R. Wallenberg, Solid-phase diffusion mechanism for GaAs nanowire growth, Nat Mater, 3 (2004) 677-681.

[123] P.L. Gai, R. Sharma, F.M. Ross, Environmental (S)TEM Studies of Gas–Liquid–Solid Interactions under Reaction Conditions, MRS Bulletin, 33 (2008) 107-

114.

[124] F.M. Ross, Controlling nanowire structures through real time growth studies, Reports on Progress in Physics, 73 (2010) 114501.

[125] F.M. Ross, In-Situ TEM Studies of Vapor- and Liquid-Phase Crystal Growth, in: In-Situ Electron Microscopy, Wiley-VCH Verlag GmbH & Co. KGaA, 2012, pp. 171-189.

[126] C.B. Maliakkal, D. Jacobsson, M. Tornberg, A.R. Persson, J. Johansson, R. Wallenberg, K.A. Dick, In situ analysis of catalyst composition during gold catalyzed GaAs nanowire growth, Nature Communications, 10 (2019) 4577.

[127] M. Tornberg, D. Jacobsson, A.R. Persson, R. Wallenberg, K.A. Dick, S. Kodambaka, Kinetics of Au-Ga Droplet Mediated Decomposition of GaAs Nanowires, Nano Lett, 19 (2019) 3498-3504.

[128] C. Hayzelden, J.L. Batstone, R.C. Cammarata, Insitu transmission electron microscopy studies of silicide-mediated crystallization of amorphous silicon, Applied Physics Letters, 60 (1992) 225-227.

[129] C. Hayzelden, J.L. Batstone, Silicide formation and silicide-mediated crystallization of nickel-implanted amorphous silicon thin films, Journal of Applied Physics, 73 (1993) 8279-8289.

[130] J.L. Batstone, In situ crystallization of amorphous silicon in the transmission electron microscope, Philosophical Magazine A, 67 (1993) 51-72.

[131] J.L. Batstone, C. Hayzelden, Microscopic Processes in Crystallisation, Solid State Phenomena, 37-38 (1994) 257-268.

[132] S. Hofmann, R. Sharma, C. Ducati, G. Du, C. Mattevi, C. Cepek, M. Cantoro, S. Pisana, A. Parvez, F. Cervantes-Sodi, A.C. Ferrari, R. Dunin-Borkowski, S. Lizzit, L. Petaccia, A. Goldoni, J. Robertson, In situ Observations of Catalyst Dynamics during Surface-Bound Carbon Nanotube Nucleation, Nano Letters, 7 (2007) 602-608.

[133] S. Kodambaka, J. Tersoff, M.C. Reuter, F.M. Ross, Germanium Nanowire Growth Below the Eutectic Temperature, Science, 316 (2007) 729-732.

[134] S. Hofmann, R. Sharma, C.T. Wirth, F. Cervantes-Sodi, C. Ducati, T. Kasama, R.E. Dunin-Borkowski, J. Drucker, P. Bennett, J. Robertson, Ledge-flow-controlled catalyst interface dynamics during Si nanowire growth, Nat Mater, 7 (2008) 372-375.

[135] B.J. Kim, J. Tersoff, S. Kodambaka, M.C. Reuter, E.A. Stach, F.M. Ross, Kinetics of Individual Nucleation Events Observed in Nanoscale Vapor-Liquid-Solid Growth, Science, 322 (2008) 1070-1073.

[136] J.-C. Harmand, G. Patriarche, F. Glas, F. Panciera, I. Florea, J.-L. Maurice, L. Travers, Y. Ollivier, Atomic Step Flow on a Nanofacet, Physical Review Letters, 121 (2018) 166101.

[137] R. Boston, Z. Schnepp, Y. Nemoto, Y. Sakka, S.R. Hall, In Situ TEM Observation of a Microcrucible Mechanism of Nanowire Growth, Science, 344 (2014) 623-626.

[138] D. Jacobsson, F. Panciera, J. Tersoff, M.C. Reuter, S. Lehmann, S. Hofmann, K.A. Dick, F.M. Ross, Interface dynamics and crystal phase switching in GaAs nanowires, Nature, 531 (2016) 317-322.

[139] A. Acrivos, THE BREAKUP OF SMALL DROPS AND BUBBLES IN SHEAR FLOWS*, Annals of the New York Academy of Sciences, 404 (1983) 1-11.

[140] J.M. Rallison, The Deformation of Small Viscous Drops and Bubbles in Shear Flows, Annual Review of Fluid Mechanics, 16 (1984) 45-66.

[141] H.A. Stone, B.J. Bentley, L.G. Leal, An experimental study of transient effects in the breakup of viscous drops, Journal of Fluid Mechanics, 173 (1986) 131-158.

[142] H.A. Stone, Dynamics of Drop Deformation and Breakup in Viscous Fluids, Annual Review of Fluid Mechanics, 26 (1994) 65-102.

[143] R. Boistelle, J.P. Astier, Crystallization mechanisms in solution, Journal of Crystal Growth, 90 (1988) 17.

[144] Z. Fan, J.L. Maurice, W. Chen, S. Guilet, E. Cambril, X. Lafosse, L. Couraud, K. Merghem, L. Yu, S. Bouchoule, I.C.P. Roca, On the Mechanism of In Nanoparticle Formation by Exposing ITO Thin Films to Hydrogen Plasmas, Langmuir, 33 (2017) 12114-12119.

[145] H.K. Yu, J.-L. Lee, Growth mechanism of metal-oxide nanowires synthesized by electron beam evaporation: A self-catalytic vapor-liquid-solid process, Sci. Rep., 4 (2014).

[146] J.O. McCaldin, T.C. McGill, The Metal-Semiconductor Interface, Annual Review of Materials Science, 10 (1980) 65-83.

[147] R. Bansen, R. Heimburger, J. Schmidtbauer, T. Teubner, T. Markurt, C. Ehlers, T. Boeck, Crystalline silicon on glass by steady-state solution growth using indium as solvent, Applied Physics A, 119 (2015) 1577-1586.

[148] C.Y. Wen, J. Tersoff, M.C. Reuter, E.A. Stach, F.M. Ross, Step-Flow Kinetics in Nanowire Growth, Physical Review Letters, 105 (2010) 195502.

[149] E. Azrak, W. Chen, S. Moldovan, S. Gao, S. Duguay, P. Pareige, P. Roca i Cabarrocas, Growth of In-Plane Ge1–xSnx Nanowires with 22 at. % Sn Using a Solid–Liquid–Solid Mechanism, The Journal of Physical Chemistry C, 122 (2018) 26236-26242.

[150] Z. Xue, M. Xu, X. Li, J. Wang, X. Jiang, X. Wei, L. Yu, Q. Chen, J. Wang,

J. Xu, Y. Shi, K. Chen, P. Roca i Cabarrocas, In-Plane Self-Turning and Twin Dynamics Renders Large Stretchability to Mono-Like Zigzag Silicon Nanowire Springs, Advanced Functional Materials, 26 (2016) 5352-5359.

[151] R.W. Olesinski, N. Kanani, G.J. Abbaschian, The In-Si (Indium-Silicon) system, Bulletin of Alloy Phase Diagrams, 6 (1985) 128-130.

[152] R.W. Olesinski, G.J. Abbaschian, The Si-Sn (Silicon-Tin) system, Bulletin of Alloy Phase Diagrams, 5 (1984) 273-276.

[153] R. Elliott, F. Shunk, The Au-Si (Gold-Silicon) system, Bulletin of Alloy Phase Diagrams, 2 (1981) 359-362.

[154] R.W. Olesinki, N. Kanani, G.J. Abbaschian, The Ge-In (Germanium-Indium) system, Bulletin of Alloy Phase Diagrams, 6 (1985) 536-539.

[155] R.W. Olesinski, N. Kanani, G.J. Abbaschian, The Ga-Si (Gallium-Silicon) system, Bulletin of Alloy Phase Diagrams, 6 (1985) 362-364.

[156] R.W. Olesinski, G.J. Abbaschian, The Ge-Sn (Germanium-Tin) system, Bulletin of Alloy Phase Diagrams, 5 (1984) 265-271.

[157] R.W. Olesinski, G.J. Abbaschian, The Pb-Si (Lead-Silicon) system, Bulletin of Alloy Phase Diagrams, 5 (1984) 271-273.

[158] P.G. Vekilov, Nucleation, Crystal Growth & Design, 10 (2010) 5007-5019.

[159] R.E. Algra, M.A. Verheijen, L.-F. Feiner, G.G.W. Immink, W.J.P.v. Enckevort, E. Vlieg, E.P.A.M. Bakkers, The Role of Surface Energies and Chemical Potential during Nanowire Growth, Nano Letters, 11 (2011) 1259-1264.

[160] B.A. Wacaser, K.A. Dick, J. Johansson, M.T. Borgström, K. Deppert, L. Samuelson, Preferential Interface Nucleation: An Expansion of the VLS Growth Mechanism for Nanowires, Advanced Materials, 21 (2009) 153-165.

[161] V.G. Dubrovskii, N.V. Sibirev, G.E. Cirlin, J.C. Harmand, V.M. Ustinov, Theoretical analysis of the vapor-liquid-solid mechanism of nanowire growth during molecular beam epitaxy, Physical Review E, 73 (2006) 021603.

[162] V.G. Dubrovskii, N.V. Sibirev, G.E. Cirlin, I.P. Soshnikov, W.H. Chen, R. Larde, E. Cadel, P. Pareige, T. Xu, B. Grandidier, J.P. Nys, D. Stievenard, M. Moewe, L.C. Chuang, C. Chang-Hasnain, Gibbs-Thomson and diffusion-induced contributions to the growth rate of Si, InP, and GaAs nanowires, Physical Review B, 79 (2009) 205316.

[163] V.G. Dubrovskii, G.E. Cirlin, N.V. Sibirev, F. Jabeen, J.C. Harmand, P. Werner, New Mode of Vapor-Liquid-Solid Nanowire Growth, Nano Letters, 11 (2011) 1247-1253.

[164] S. Kodambaka, J. Tersoff, M.C. Reuter, F.M. Ross, Diameter-Independent Kinetics in the Vapor-Liquid-Solid Growth of Si Nanowires, Physical Review Letters, 96 (2006) 096105.

[165] D.J. Chadi, Theoretical study of the atomic structure of silicon (211), (311),

and (331) surfaces, Physical Review B, 29 (1984) 785-792.

[166] A.A. Baski, L.J. Whitman, S.C. Erwin, A Stable High-Index Surface of Silicon: Si(5 5 12), Science, 269 (1995) 1556-1560.

[167] A.A. Stekolnikov, J. Furthmüller, F. Bechstedt, Absolute surface energies of group-IV semiconductors: Dependence on orientation and reconstruction, Physical Review B, 65 (2002) 115318.

[168] E.I. Givargizov, Fundamental aspects of VLS growth, Journal of Crystal Growth, 31 (1975) 20-30.

[169] K.W. Schwarz, J. Tersoff, From Droplets to Nanowires: Dynamics of Vapor-Liquid-Solid Growth, Physical Review Letters, 102 (2009) 206101.

[170] K.W. Schwarz, J. Tersoff, Elementary Processes in Nanowire Growth, Nano Letters, 11 (2011) 316-320.

[171] K.W. Schwarz, J. Tersoff, Multiplicity of Steady Modes of Nanowire Growth, Nano Letters, 12 (2012) 1329-1332.

[172] K.W. Schwarz, J. Tersoff, S. Kodambaka, F.M. Ross, Jumping-Catalyst Dynamics in Nanowire Growth, Phys. Rev. Lett., 113 (2014).

[173] J. Kočka, M. Müller, J. Stuchlík, H. Stuchlíková, J. Červenka, A. Fejfar, Role of a-Si:H in lateral growth of crystalline silicon nanowires using Pb and In catalysts, physica status solidi (a), (2016) n/a-n/a.

[174] M.S. Seifner, A. Dijkstra, J. Bernardi, A. Steiger-Thirsfeld, M. Sistani, A. Lugstein, J.E.M. Haverkort, S. Barth, Epitaxial Ge0.81Sn0.19 Nanowires for Nanoscale Mid-Infrared Emitters, ACS Nano, 13 (2019) 8047-8054.

[175] Y. Shan, A.K. Kalkan, C.-Y. Peng, S.J. Fonash, From Si Source Gas Directly to Positioned, Electrically Contacted Si Nanowires: The Self-Assembling Grow-in-Place Approach, Nano Letters, 4 (2004) 2085-2089.

[176] Y. Shan, S.J. Fonash, Self-Assembling Silicon Nanowires for Device Applications Using the Nanochannel-Guided "Grow-in-Place" Approach, ACS Nano, 2 (2008) 429-434.

[177] A. Ismach, L. Segev, E. Wachtel, E. Joselevich, Atomic-Step-Templated Formation of Single Wall Carbon Nanotube Patterns, Angewandte Chemie, 116 (2004) 6266-6269.

[178] A. Ismach, D. Kantorovich, E. Joselevich, Carbon Nanotube Graphoepitaxy: Highly Oriented Growth by Faceted Nanosteps, Journal of the American Chemical Society, 127 (2005) 11554-11555.

[179] D. Tsivion, M. Schvartzman, R. Popovitz-Biro, P. von Huth, E. Joselevich, Guided Growth of Millimeter-Long Horizontal Nanowires with Controlled Orientations, Science, 333 (2011) 1003-1007.

[180] D. Tsivion, M. Schvartzman, R. Popovitz-Biro, E. Joselevich, Guided Growth of Horizontal ZnO Nanowires with Controlled Orientations on Flat and

Faceted Sapphire Surfaces, ACS Nano, 6 (2012) 6433-6445.

[181] M. Schvartzman, D. Tsivion, D. Mahalu, O. Raslin, E. Joselevich, Self-integration of nanowires into circuits via guided growth, PNAS, (2013).

[182] D. Tsivion, E. Joselevich, Guided Growth of Epitaxially Coherent GaN Nanowires on SiC, Nano Letters, 13 (2013) 5491-5496.

[183] D. Tsivion, E. Joselevich, Guided Growth of Horizontal GaN Nanowires on Spinel with Orientation-Controlled Morphologies, The Journal of Physical Chemistry C, 118 (2014) 19158-19164.

[184] E. Oksenberg, R. Popovitz-Biro, K. Rechav, E. Joselevich, Guided Growth of Horizontal ZnSe Nanowires and their Integration into High-Performance Blue-UV Photodetectors, Adv Mater, 27 (2015) 3999-4005.

[185] B. Nikoobakht, C.A. Michaels, S.J. Stranick, M.D. Vaudin, Horizontal growth and in situ assembly of oriented zinc oxide nanowires, Applied Physics Letters, 85 (2004) 3244-3246.

[186] S. Han, X. Liu, C. Zhou, Template-Free Directional Growth of Single-Walled Carbon Nanotubes on a- and r-Plane Sapphire, Journal of the American Chemical Society, 127 (2005) 5294-5295.

[187] P.-J. Alet, L. Yu, G. Patriarche, S. Palacin, P. Roca i Cabarrocas, In situ generation of indium catalysts to grow crystalline silicon nanowires at low temperature on ITO, Journal of Materials Chemistry, 18 (2008) 5187-5189.

[188] N. Hideo, I. Ryuichi, H. Yosuke, S. Hiroyuki, Effects of surface roughness on wettability, Acta Materialia, 46 (1998) 2313-2318.

[189] S.A. Fortuna, J. Wen, I.S. Chun, X. Li, Planar GaAs Nanowires on GaAs (100) Substrates: Self-Aligned, Nearly Twin-Defect Free, and Transfer-Printable, Nano Letters, 8 (2008) 4421-4427.

[190] S.A. Fortuna, X. Li, Metal-catalyzed semiconductor nanowires: a review on the control of growth directions, Semiconductor Science and Technology, 25 (2010) 024005.

[191] C. Zhang, X. Miao, P.K. Mohseni, W. Choi, X. Li, Site-Controlled VLS Growth of Planar Nanowires: Yield and Mechanism, Nano Letters, 14 (2014) 6836-6841.

[192] P. Tejedor, P. Šmilauer, C. Roberts, B.A. Joyce, Surface-morphology evolution during unstable homoepitaxial growth of GaAs(110), Physical Review B, 59 (1999) 2341-2345.

[193] J. Sudijono, M.D. Johnson, C.W. Snyder, M.B. Elowitz, B.G. Orr, Surface evolution during molecular-beam epitaxy deposition of GaAs, Physical Review Letters, 69 (1992) 2811-2814.

[194] M.D. Johnson, J. Sudijono, A.W. Hunt, B.G. Orr, Growth mode evolution during homoepitaxy of GaAs (001), Applied Physics Letters, 64 (1994) 484-486.

[195] L. Yu, M. Oudwan, O. Moustapha, F. Fortuna, P. Roca i Cabarrocas, Guided growth of in-plane silicon nanowires, Applied Physics Letters, 95 (2009) 113106-113103.

[196] L. Yu, W. Chen, B. O'Donnell, G. Patriarche, S. Bouchoule, P. Pareige, R. Rogel, A.C. Salaun, L. Pichon, P. Roca i Cabarrocas, Growth-in-place deployment of in-plane silicon nanowires, Applied Physics Letters, 99 (2011) 203104-203103.

[197] M. Xu, Z. Xue, L. Yu, S. Qian, Z. Fan, J. Wang, J. Xu, Y. Shi, K. Chen, P. Roca i Cabarrocas, Operating principles of in-plane silicon nanowires at simple step-edges, Nanoscale, 7 (2015) 5197-5202.

[198] M. Xu, J. Wang, Z. Xue, J. Wang, P. Feng, L. Yu, J. Xu, Y. Shi, K. Chen, I.C.P. Roca, High performance transparent in-plane silicon nanowire Fin-TFTs via a robust nano-droplet-scanning crystallization dynamics, Nanoscale, 9 (2017) 10350-10357.

[199] Z. Xue, T. Dong, Z. Zhu, Y. Zhao, Y. Sun, L. Yu, Engineering in-plane silicon nanowire springs for highly stretchable electronics, Journal of Semiconductors, 39 (2018) 011001.

[200] D. Turnbull, Kinetics of Heterogeneous Nucleation, The Journal of Chemical Physics, 18 (1950) 198-203.

[201] T.P. Melia, Crystal nucleation from aqueous solution, Journal of Applied Chemistry, 15 (1965) 345-357.

[202] C.C. Tsai, J.C. Knights, G. Chang, B. Wacker, Film formation mechanisms in the plasma deposition of hydrogenated amorphous silicon, Journal of Applied Physics, 59 (1986) 2998-3001.

[203] J.C. Knights, R.A. Lujan, Microstructure of plasma‐deposited a‐Si : H films, Applied Physics Letters, 35 (1979) 244-246.

[204] J.B. Hannon, S. Kodambaka, F.M. Ross, R.M. Tromp, The influence of the surface migration of gold on the growth of silicon nanowires, Nature, 440 (2006) 69-71.

[205] M.I. den Hertog, J.-L. Rouviere, F. Dhalluin, P.J. Desré, P. Gentile, P. Ferret, F. Oehler, T. Baron, Control of Gold Surface Diffusion on Si Nanowires, Nano Letters, 8 (2008) 1544-1550.

[206] T. Kawashima, T. Mizutani, T. Nakagawa, H. Torii, T. Saitoh, K. Komori, M. Fujii, Control of Surface Migration of Gold Particles on Si Nanowires, Nano Letters, 8 (2008) 362-368.

[207] S. Kodambaka, J.B. Hannon, R.M. Tromp, F.M. Ross, Control of Si Nanowire Growth by Oxygen, Nano Letters, 6 (2006) 1292-1296.

[208] W. Chen, P. Roca i Cabarrocas, Insights into gold-catalyzed plasma-assisted CVD growth of silicon nanowires, Applied Physics Letters, 109 (2016) 043108.

[209] I. Zardo, S. Conesa-Boj, S. Estradé, L. Yu, F. Peiro, P. Roca i Cabarrocas, J.R. Morante, J. Arbiol, A. Fontcuberta i Morral, Growth study of indium-catalyzed silicon nanowires by plasma enhanced chemical vapor deposition, Applied Physics A, 100 (2010) 287-296.

[210] Y. Linwei, A. Pierre-Jean, P. Gennaro, M. Isabelle, C. Pere Roca i, Synthesis, morphology and compositional evolution of silicon nanowires directly grown on SnO 2 substrates, Nanotechnology, 19 (2008) 485605.

[211] S. Misra, L. Yu, W. Chen, P. Roca i Cabarrocas, Wetting Layer: The Key Player in Plasma-Assisted Silicon Nanowire Growth Mediated by Tin, The Journal of Physical Chemistry C, 117 (2013) 17786-17790.

[212] W. Chen, L. Yu, S. Misra, Z. Fan, P. Pareige, G. Patriarche, S. Bouchoule, P.R.i. Cabarrocas, Incorporation and redistribution of impurities into silicon nanowires during metal-particle-assisted growth, Nat Commun, 5 (2014).

[213] B. Drevillon, J. Perrin, J.M. Siefert, J. Huc, A. Lloret, G. de Rosny, J.P.M. Schmitt, Growth of hydrogenated amorphous silicon due to controlled ion bombardment from a pure silane plasma, Applied Physics Letters, 42 (1983) 801-803.

[214] J. Perrin, P.R.i. Cabarrocas, B. Allain, J.-M. Friedt, a-Si:H Deposition from SiH 4 and Si 2 H 6 rf-Discharges: Pressure and Temperature Dependence of Film Growth in Relation to α-γ Discharge Transition, Japanese Journal of Applied Physics, 27 (1988) 2041.

[215] J. Perrin, Plasma and surface reactions during a-Si:H film growth, Journal of Non-Crystalline Solids, 137–138, Part 2 (1991) 639-644.

[216] A. Fontcuberta i Morral, P. Roca i Cabarrocas, Role of hydrogen diffusion on the growth of polymorphous and microcrystalline silicon thin films, The European Physical Journal - Applied Physics, 35 (2006) 165-172.

[217] A.M. Antoine, B. Drevillon, P. Roca, I. Cabarrocas, Growth processes of RF glow discharge deposited a-Si:H and a-Ge:H films, Journal of Non-Crystalline Solids, 77–78, Part 2 (1985) 769-772.

[218] A.M. Antoine, B. Drevillon, P. Roca i Cabarrocas, Insitu investigation of the growth of rf glow-discharge deposited amorphous germanium and silicon films, Journal of Applied Physics, 61 (1987) 2501-2508.

[219] P. Roca i Cabarrocas, J.B. Chévrier, J. Huc, A. Lloret, J.Y. Parey, J.P.M. Schmitt, A fully automated hot-wall multiplasma-monochamber reactor for thin film deposition, Journal of Vacuum Science & Technology A, 9 (1991) 2331-2341.

[220] C. Godet, N. Layadi, P. Roca i Cabarrocas, Role of mobile hydrogen in the amorphous silicon recrystallization, Applied Physics Letters, 66 (1995) 3146-3148.

[221] E.D. Leshchenko, P. Kuyanov, R.R. LaPierre, V.G. Dubrovskii, Tuning the morphology of self-assisted GaP nanowires, Nanotechnology, 29 (2018) 225603.

[222] J. Tersoff, Stable Self-Catalyzed Growth of III-V Nanowires, Nano Lett, 15

(2015) 6609-6613.

[223] M. Foldyna, L. Yu, P. Roca i Cabarrocas, Theoretical short-circuit current density for different geometries and organizations of silicon nanowires in solar cells, Solar Energy Materials and Solar Cells, 117 (2013) 645-651.

[224] D.E. Perea, E.R. Hemesath, E.J. Schwalbach, J.L. Lensch-Falk, P.W. Voorhees, L.J. Lauhon, Direct measurement of dopant distribution in an individual vapour-liquid-solid nanowire, Nat Nano, 4 (2009) 315-319.

[225] P.V. Radovanovic, Nanowires: Keeping track of dopants, Nat Nano, 4 (2009) 282-283.

[226] V.G. Dubrovskii, Influence of the group V element on the chemical potential and crystal structure of Au-catalyzed III-V nanowires, Applied Physics Letters, 104 (2014) 053110.

[227] L.C. Campos, V.R. Manfrinato, J.D. Sanchez-Yamagishi, J. Kong, P. Jarillo-Herrero, Anisotropic Etching and Nanoribbon Formation in Single-Layer Graphene, Nano Letters, 9 (2009) 2600-2604.

[228] L. Ci, Z. Xu, L. Wang, W. Gao, F. Ding, K.F. Kelly, B.I. Yakobson, P.M. Ajayan, Controlled nanocutting of graphene, Nano Research, 1 (2008) 116-122.

[229] S. Li, Y.-C. Lin, W. Zhao, J. Wu, Z. Wang, Z. Hu, Y. Shen, D.-M. Tang, J. Wang, Q. Zhang, H. Zhu, L. Chu, W. Zhao, C. Liu, Z. Sun, T. Taniguchi, M. Osada, W. Chen, Q.-H. Xu, A.T.S. Wee, K. Suenaga, F. Ding, G. Eda, Vapour–liquid–solid growth of monolayer MoS2 nanoribbons, Nature Materials, 17 (2018) 535-542.